\documentclass[amsmath,aps,prd,twocolumn,showpacs,showkeys,preprintnumbers,bibnotes,floatfix,longbibliography,notitlepage,nofootinbib,superscriptaddress,superscriptgaddress
]{revtex4-2}

\pdfoutput=1
\usepackage{amsmath}
\usepackage{amsfonts}
\usepackage{amssymb}
\usepackage{mathrsfs}
\usepackage{graphicx}
\usepackage{color}
\usepackage{bbding}
\usepackage{tabularx}
\usepackage[usenames,dvipsnames,table]{xcolor}
\usepackage[normalem]{ulem}
\usepackage{makecell}
\usepackage{slashed}

\usepackage[hidelinks]{hyperref}
\usepackage{multirow}
\usepackage{makecell}
\usepackage{upgreek}
\usepackage[capitalise]{cleveref}

\newcommand{\meson}{\mathfrak{m}}

\newcommand{\beq}{\begin{equation}}
\newcommand{\eeq}{\end{equation}}

\begin{document}

\title{Searching for heavy millicharged particles from the atmosphere}

\author{Han Wu}
\email{han.wu@queensu.ca}
\affiliation{Department of Physics, Engineering Physics and Astronomy, Queen's University, Kingston, Ontario, K7L 3N6, Canada}
\affiliation{Arthur B. McDonald Canadian Astroparticle Physics Research Institute, Kingston, Ontario, K7L 3N6, Canada}

\author{Edward Hardy}
\email{edward.hardy@physics.ox.ac.uk}
\affiliation{
Rudolf Peierls Centre for Theoretical Physics, University of Oxford, OX1 3PU, Parks Road, Oxford, United Kingdom}

\author{Ningqiang Song}
\email{songnq@itp.ac.cn}
\affiliation{Institute of Theoretical Physics, Chinese Academy of Sciences, Beijing, 100190, China}
\affiliation{Department of Mathematical Sciences, University of Liverpool, Liverpool, L69 7ZL, United Kingdom}

\date{\today}

\begin{abstract}
If millicharged particles (MCPs) exist they can be created in the atmosphere when high energy cosmic rays collide with nuclei and could subsequently be detected at neutrino experiments. We extend previous work, which considered MCPs from decays of light mesons and proton bremsstrahlung, by including production from $\Upsilon$ meson decays and the Drell-Yan process. MCPs with masses below a GeV primarily arise from proton bremsstrahlung, while heavier MCPs predominantly originate from heavy meson decays and Drell-Yan. We analyse the resulting single scatter and multiple scatter signals at SuperK and JUNO. Searches for low energy coincident signals at JUNO will be sensitive to MCPs with milli-charges up to an order of magnitude beyond current constraints for MCP masses between 2 GeV and 10 GeV.
\end{abstract}

\maketitle

\section{Introduction}

The existence of dark matter, along with current understanding of the richness of typical string theory compactifications \cite{Abel:2008ai,Abel:2006qt,Acharya:2018deu}, motivates the presence of additional ``dark" sectors that are only very weakly coupled to the Standard Model (SM). Among the new degrees of freedom that might reside within such a sector, millicharged particles (i.e. particles with a very small charge $\epsilon e$ under electromagnetism) are a particularly interesting possibility, being both theoretically motivated and leading to new experimental and observational signals. 

Millicharged particles (MCPs) arise naturally from one of the most minimal portals between a dark sector and the SM: kinetic mixing of a new U(1) gauge boson (a ``dark photon'') and the photon of the form $\frac{1}{2}\epsilon F F'$, where $F$ and $F'$ are the electromagnetic and dark U(1) field strengths respectively~\cite{Holdom:1985ag,Dienes:1996zr,Abel:2003ue}. Indeed, even if such a mixing is absent at high energy scales, if there are heavy particles charged under both the new $U(1)$ and the SM hypercharge (with charges $q'_i$, $q_{i}$ with respect to the two groups and masses $M_{i}$) then a one-loop diagram leads to a kinetic mixing at low scales of
\beq
\epsilon = \frac{-g  g'}{16\pi^2} \sum\limits_i q'_iq_{i}\ln\left(\frac{M_{i}^2}{\mu^2} \right)~,
\label{eq:kineticloop}
\eeq
where $g'$, $g$ are the gauge couplings of the two U(1)s and  $\mu$ is the renormalization scale \cite{Holdom:1985ag,Cheung:2009qd,Gherghetta:2019coi}. Eq.~\eqref{eq:kineticloop} highlights  $\epsilon\in(10^{-3},10^{-1})$ as an especially plausible range if $g'$ is not too much smaller than the SM gauge couplings. If the dark photon is massless, the kinetic mixing can be rotated away through field redefinitions such that the dark photon only couples to dark sector particles, while matter charged under the dark U(1) gets a millicharge of approximately $\epsilon e$~\cite{Fabbrichesi:2020wbt}. Additionally, millicharged particles (MCPs) might provide viable explanations for the EDGES anomaly~\cite{Bowman:2018yin,Munoz:2018pzp,Berlin:2018sjs,Kovetz:2018zan,Liu:2019knx,Aboubrahim:2021ohe,Mathur:2021gej} and possible deviations in the muon magnetic moment~\cite{Bai:2021nai} thanks to their coupling to photons, and they might comprise dark matter.

MCPs have been a key target of many experimental searches for physics beyond the SM. Sources of current bounds include electron~\cite{Prinz:1998ua,Berlin:2018bsc,Gninenko:2018ter,Chu:2018qrm} and proton~\cite{Golowich:1986tj,Soper:2014ska,Magill:2018tbb,Harnik:2019zee} beam dumps, and collider searches~\cite{Davidson:1991si,Davidson:2000hf,Liu:2018jdi,Ball:2020dnx}. Various new experiments are also planned~\cite{Kelly:2018brz,Liu:2019ogn,Foroughi-Abari:2020qar,Liang:2019zkb,Gorbunov:2021jog,Carney:2021irt,Budker:2021quh,Gorbunov:2022bzi,Kling:2022ykt,Mitsou:2023dkg}. However, despite these extensive efforts, large parts of MCP parameter space remain unexplored even for $\epsilon\gtrsim 10^{-3}$, especially for MCP masses around or above a GeV.

\begin{figure*}[!t]
\centering
\includegraphics[width=1.41\columnwidth ]{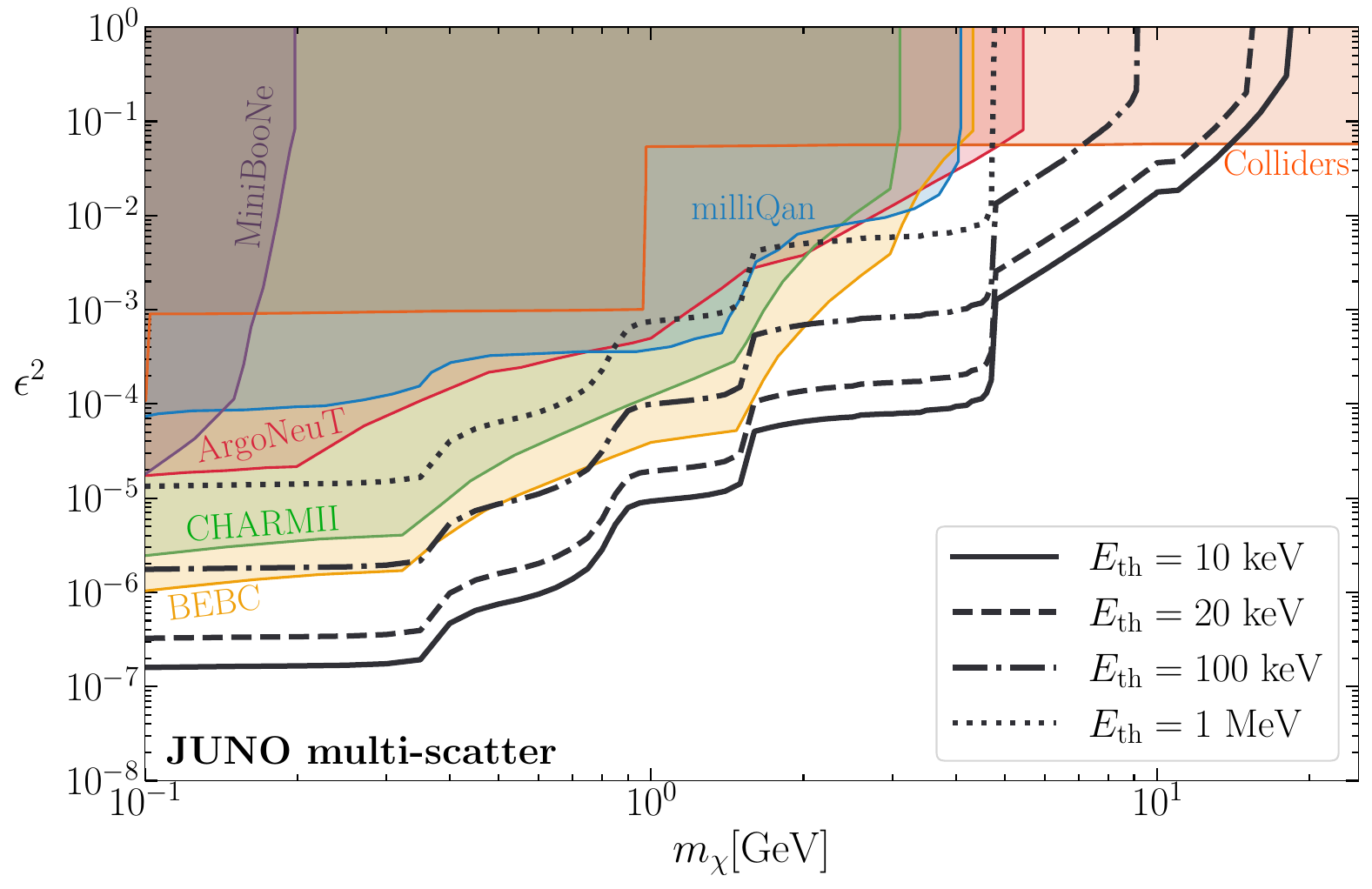}
\caption{Future sensitivity of JUNO to millicharged particles of charge $\epsilon e$ via searches for multiple scatter events (black lines), accounting for production in the atmosphere by meson decay, proton bremsstrahlung and Drell-Yan. Results are obtained assuming an exposure of 170 kton$\cdot$yr and are plotted for different detector energy thresholds $E_{\rm th}$. Overlaid are existing constraints from BEBC and CHARM II~\cite{marocco2021blast}, MiniBooNE~\cite{aguilar2018dark}, ArgoNeuT~\cite{acciarri2020improved}, milliQan~\cite{lowette2021sensitivity} and colliders~\cite{davidson2000updated}  (some of these might not apply at large $\epsilon^2\gtrsim 10^{-1}$ due to the millicharged particles scattering prior to reaching the detector).}
\label{fig:results_multi}
\end{figure*} 

A recently suggested approach to discovering new weakly-coupled physics is to make use of the cosmic rays that bombard the atmosphere~\cite{Plestid:2020kdm,Kachelriess:2021man,ArguellesDelgado:2021lek,Alvey:2019zaa,Su:2020zny}. In particular, collisions between cosmic rays and nuclei in the atmosphere can produce beyond-SM particles, which can be searched for at neutrino or dark matter experiments. 
The flux of MCPs from the decay of light mesons produced by such collisions has been studied in detail in Refs.~\cite{Plestid:2020kdm,Kachelriess:2021man,ArguellesDelgado:2021lek}, and the flux from bremsstrahlung from an incoming cosmic ray proton has been analysed in Ref.~\cite{Du:2022hms}. The resulting constraints (from single scatter events) are competitive with other limits on MCPs and there is possible strong future sensitivity (from multiple scatter search strategies) to MCPs with masses between $100~{\rm MeV}$ and a few ${\rm GeV}$.

In this paper, we study the production of relatively heavy MCPs in the atmosphere through $\Upsilon$ meson decay and Drell-Yan. 
We analyze the detection of such particles at the upcoming JUNO experiment, which will have a trigger system that allows signals from a single MCP scattering multiple times to be identified. As summarised in Figure~\ref{fig:results_multi}, despite the MCP flux being relatively small, the low threshold energies and large exposure at JUNO will allow a substantial part of currently-unexplored parameter space in the mass range 1-10~GeV to be covered, including values of $\epsilon$ motivated by Eq.~\eqref{eq:kineticloop}. We also include the production of MCPs by proton bremsstrahlung making use of the Fermi-Weizacker-Williams approximation, which yields conservative results.

The remainder of this paper is structured as follows. We compute the MCP flux from meson decay, proton bremsstrahlung and the Drell-Yan process in Section~\ref{sec:MCPproduction}. In  Section~\ref{sec:detection} we describe the propagation and detection of MCPs. We present our main results in Section~\ref{sec:results} and conclude in Section~\ref{sec:discussions}.

\section{Millicharged particle production}
\label{sec:MCPproduction}

Throughout, we consider the SM extended by a single millicharged Dirac fermion $\chi$ with mass $m_\chi$ and electromagnetic charge $\epsilon e$. As we will see, the dominant production mechanism in the atmosphere depends on $m_\chi$.

\subsection{Meson decay}

High energy cosmic rays undergo inelastic collisions with nuclei in the atmosphere, generating a cascade of particles, including photons, leptons and hadrons. As first pointed out and analysed in Ref.~\cite{Plestid:2020kdm},  mesons (which we denote $\mathfrak{m}$) are particularly important for MCP production. As usual, these can be grouped into pseudoscalar mesons $\pi^0$ and $\eta$, and vector mesons $\rho^0$, $\omega$, $\phi$, $J/\psi$ and $\Upsilon$ according to their spins and parities. Pseudoscalar mesons can decay into MCPs via the process $\mathfrak{m}\rightarrow \chi\bar{\chi}\gamma$, while vector mesons decay into a pair of MCPs directly: $\mathfrak{m}\rightarrow \chi\bar{\chi}$. 
We do not consider MCP production from pions because this is only possible for MCPs with mass $m_{\chi}\lesssim 60~{\rm MeV}$, which are already severely constrained by terrestrial experiments, in particular the MilliQ experiment at SLAC \cite{Prinz:1998ua}.\footnote{Production from kaons is negligible because, for fixed $\epsilon^2$, their branching fraction to MCPs is much smaller than other mesons.}
We summarize the calculation of the resulting MCP flux below; further details of the cosmic-ray proton flux, meson production cross section and the decay branching ratios and kinematics can be found in Appendix~\ref{app:mesondecay}.

We calculate the meson flux semi-analytically, following Ref.~\cite{Plestid:2020kdm}. The differential flux of mesons   $\Phi_\meson\equiv d^2N/(dE_\meson d\Omega)$ is obtained from the fraction of cosmic-ray protons that interact to produce a meson of energy $E_\meson$ 
(making the excellent approximation that all cosmic-ray protons scatter in the atmosphere):
\begin{equation}
    \Phi_\meson(E_\meson)=\int  dE_p~\Phi_p(E_p)\frac{1}{\sigma_{pp}(E_p)}\dfrac{d\sigma_{pp}^\meson(E_p)}{dE_\meson}\,,
    \label{eq:dphidEm}
\end{equation}
where $\Phi_p$ is the cosmic ray proton flux with $E_p$ the energy of the incoming proton, $d\sigma_{pp}^\meson/dE_\meson$ is the differential cross section for producing mesons in the process $pp\rightarrow \meson+ X$ and $\sigma_{pp}$ is the total cross section for $pp$ collisions, which is dominated by inelastic scattering in the relevant energy range~\cite{particle2022review}.\footnote{$\sigma_{pp}$ and $\sigma_{pp}^\meson$ should actually be the corresponding cross sections for cosmic rays scattering with atmospheric nuclei. However, both cross sections are expected to scale the same way with the number of nucleons in a nucleus, so we can use the (better-known) cross sections for $pp$ collisions.} 

We use the analytic model of the cosmic ray flux implemented in DarkSUSY~\cite{bringmann2018darksusy}, which fits observational data well and decreases approximately proportionally to $E_p^{-2.7}$ for the $E_p$ of interest (see Ref.~\cite{boschini2017solution} for details). 
Additionally, in Eq.~\eqref{eq:dphidEm} we assume that mesons are only produced in the first collision of a cosmic ray, with the cosmic-ray protons subsequently lost regardless of the type of interaction. Although conservative compared to the alternative of calculating the meson flux numerically by solving the cascade equations (e.g. as in Refs.~\cite{Fedynitch:2015zma,Fedynitch:2012fs}), this is a reasonable approximation given the dominance of inelastic scattering. The difference in the $\Phi_\meson$ obtained from the two approaches is typically less than a factor of two, leading to only a minor difference in the experimental sensitivity to MCPs.

The cross sections $d\sigma_{pp}^\meson/dE_\meson$ for production of relatively light mesons (in particular the $\eta$, $\rho$, $\omega$, $\phi$ and $J/\psi$, as considered in the previous literature) can be conveniently extracted from experimental data \cite{Plestid:2020kdm}. 
In addition, we include MCPs produced from $\Upsilon$ mesons. The large $\Upsilon$ mass (approximately $9.46$~GeV) allows relatively heavy MCPs, with masses up to about $4.7~{\rm GeV}$, to be produced. Although a few measurements of $\Upsilon$ production have been made at colliders at different beam energies~\cite{CMS:2020qwa,ATLAS:2012lmu,STAR:2010wzc}, due to kinematic cuts the total cross section is not straightforward to determine. Instead, we calculate the $\Upsilon$ flux using \texttt{Pythia 8.3}~\cite{Bierlich:2022pfr}. We focus on the contribution from the $\Upsilon 1S$ state and have checked that the production of higher resonances is strongly suppressed, which is consistent with the discussion in Ref.~\cite{Foroughi-Abari:2020qar}.

We plot the resulting meson flux in the left panel of Figure~\ref{fig:MCPproduction} in Appendix~\ref{app:mesondecay}. Low mass mesons are readily produced from cosmic ray collisions, while the total rate of production of heavier mesons is strongly suppressed. This is due to the scarcity of sufficiently energetic cosmic ray protons and because, at a given center of mass energy, the cross section for producing heavy mesons is substantially smaller than for light mesons.

Having obtained the flux of mesons, the differential flux of MCPs of energy $E_\chi$ originating from meson decay can be computed as
\begin{equation} \label{eq:Phichi}
\Phi_\chi^{\meson}\left(E_\chi\right)=2\sum_{\meson}\textrm{Br}\left(\meson\rightarrow\chi\bar{\chi} \left(\gamma\right) \right)\int dE_\meson~\Phi_\meson\dfrac{1}{\Gamma_\meson}\dfrac{d\Gamma_\meson}{dE_\chi}\,,
\end{equation}
where $\textrm{Br}(\meson\rightarrow\chi\bar\chi (\gamma))$ is the branching ratio for meson decay into a pair of MCPs, and $d\Gamma_\meson/dE_\chi$ is the energy distribution of MCPs in the decays  (with $\Gamma_\meson$ the total decay rate of the meson to MCPs). 
We obtain the meson branching ratio to MCPs from the measured branching ratio of meson decays to a pair of muons, relative to which MCP production is suppressed by a factor of $\epsilon^2$,  along with a difference from the phase space. 
Not surprisingly, a MCP of mass $m_\chi$ is dominantly produced by the lightest meson with mass greater than $2m_\chi$.

\subsection{Proton bremsstrahlung}

\begin{figure}[!t]
\centering
\includegraphics[width=0.49\columnwidth]{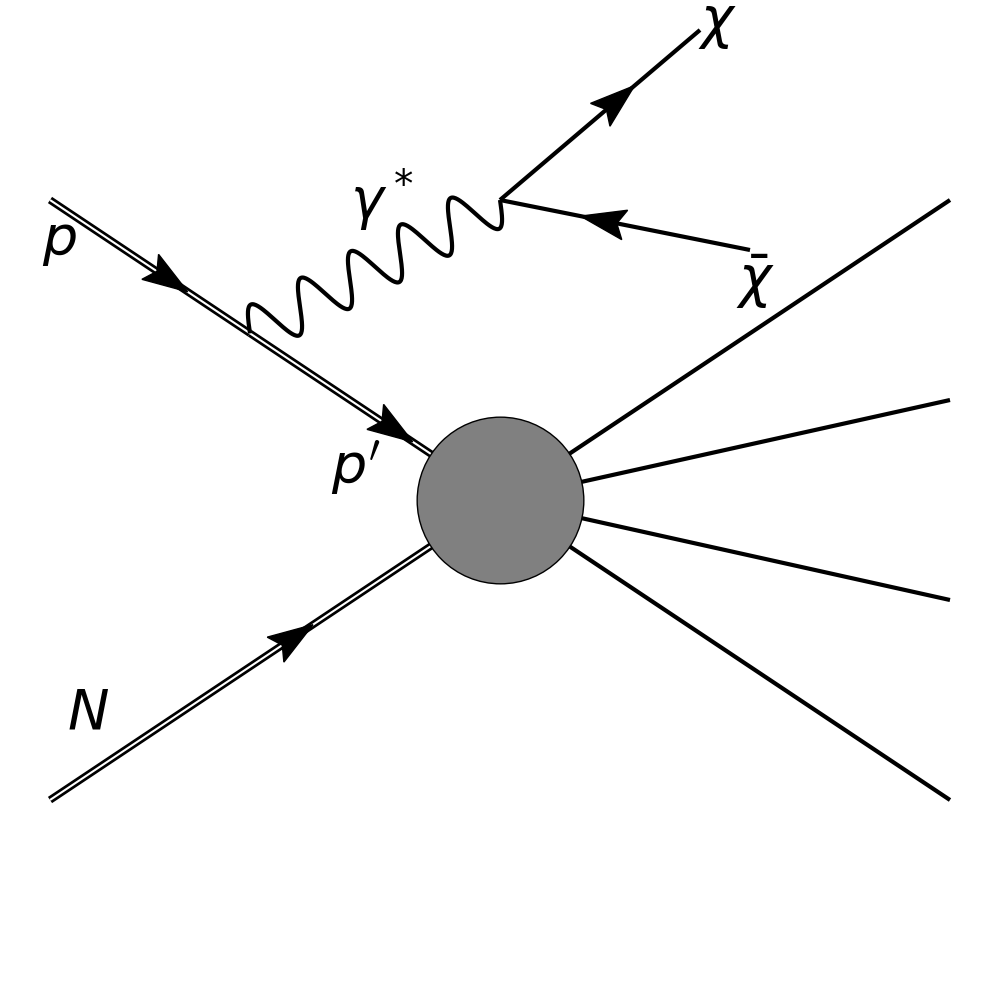}
\includegraphics[width=0.49\columnwidth]{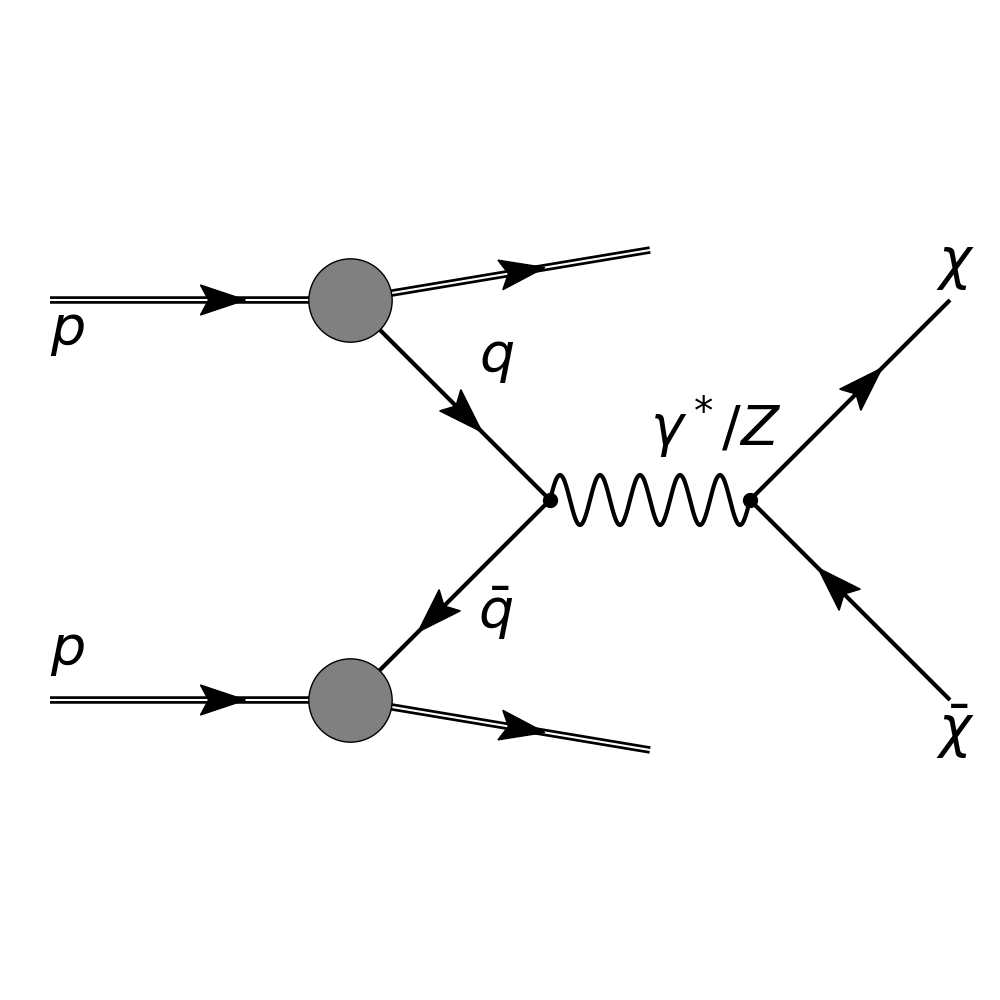}
\caption{Feynman diagrams for MCP production through proton bremsstrahlung (left) and the Drell-Yan process (right).}
\label{fig:feynman}
\end{figure}

MCPs can also be produced via bremsstrahlung~\cite{deNiverville:2016rqh}, with a virtual photon emitted by the incoming cosmic ray proton converting to a pair of MCPs as shown in the Feynman diagram in Figure~\ref{fig:feynman} left. In the context of beam dump experiments,  the rate at which this process occurs has been evaluated using the Fermi-Weizs\"acker-Williams (FWW) approximation within the splitting-kernel approach~\cite{Blumlein:2013cua,deNiverville:2016rqh,Tsai:2019buq,Feng:2017uoz,Foroughi-Abari:2021zbm,Du:2021cmt}. Recently, Ref.~\cite{Du:2022hms} applied the splitting-kernel method to compute the atmospheric MCP flux from proton bremsstrahlung, considering both the FWW approximation and a new way of evaluating the splitting-kernel. 
We adopt the conventional FWW approach, which, as discussed in Ref.~\cite{Du:2022hms}, yields conservative results. In particular, in the FWW approach we impose kinematic cuts such that the required relativistic and colinear conditions are satisfied, which result in a total cross section for MCP production that is smaller that using the method of Ref.~\cite{Du:2022hms}, see Appendix~\ref{app:protonbrem} for details.

Given the composition of the atmosphere, the relevant process is bremsstrahlung when cosmic ray protons scatter with nitrogen nuclei. 
We consider only production from interactions in which the cosmic rays are ultra-relativistic and the initial state radiation is close to collinear, because a reliable calculation of the MCP flux is possible in this regime. In such a limit the intermediate proton is nearly on-shell, and the cross section for proton nitrogen scattering with bremsstrahlung from the initial proton  $\sigma_{pN}^{\rm PB}$ can be matched to the proton-nitrogen cross section $\sigma_{pN}$ via
\begin{equation}
    d\sigma^{\rm PB}_{pN}(s,k)\simeq d\mathcal{P}_{p\rightarrow \gamma^*p^\prime}\times \sigma_{pN}(s')\,,
    \label{eq:dsigmaPB}
\end{equation}
where $k$ is the four momentum of the virtual photon, $s=(p+p_N)^2$, $s'=(p+p_N-k)^2$, and $p$ and $p_N$ are the four-momenta of the incoming proton and nitrogen nucleus \cite{Blumlein:2013cua,deNiverville:2016rqh,Foroughi-Abari:2021zbm,Gorbunov:2023jnx,Du:2022hms} (see e.g. Appendix C of~\cite{Foroughi-Abari:2021zbm} for a derivation). 

In Eq.~\eqref{eq:dsigmaPB}, $d\mathcal{P}_{p\rightarrow \gamma^*p^\prime}$  represents the probability for the proton-bremsstrahlung sub-process $p\rightarrow \gamma^*p^\prime$ to occur, where $p'= p-k$. 
In the ultra-relativistic and collinear limit, $d\mathcal{P}_{p\rightarrow \gamma^*p^\prime}$ can be evaluated using the FWW approximation, in which the $pp$ scattering is assumed to be dominated by the exchange of a boson mediator with small virtuality. 
We also include the electromagnetic form factor of the proton in computing the radiation probability, detailed in Appendix~\ref{app:protonbrem}, which can be described phenomenologically as a resonant enhancement due to mixing between vector mesons and the virtual photon \cite{Faessler:2009tn,deNiverville:2016rqh} (this was also included in the analysis of Ref.~\cite{Du:2022hms}). When the momentum of the virtual photon is larger than a GeV, this form factor tends to sharply suppress the emission probability. 
The form factor is a phenomenological description of the proton's electromagnetic properties using the vector meson dominance model~\cite{Faessler:2009tn}. We note there is double counting with meson decay for the part of the form factor when the meson mediator is produced on-shell in proton bremsstrahlung. However, as we will show later (see in particular Figure~\ref{fig:prod}), in the regime $m_\chi\lesssim$~GeV where the form factor is effective, the MCP flux is always dominated by proton bremsstrahlung. Meanwhile for larger MCP mass, MCPs dominantly originate from heavy meson decay (and Drell-Yan described in the next subsection), which is not captured by the form factor.As a result, the sensitivity that we would obtain by, for each MCP mass, considering only the dominant production process (which prevents any double-counting) is for practical purposes the same as the results we obtain, as can be inferred from Figure~\ref{fig:prod}.

The differential flux of MCPs from proton bremsstrahlung is then given by~\cite{Gninenko:2018ter,Du:2022hms}
\begin{equation}
\begin{split}
    \Phi_\chi^{\rm PB}\left(E_\chi\right) &=\int dE_p~\Phi_p(E_p)\dfrac{\epsilon^2 e^2}{6\pi^2} \\
    &\times\int \dfrac{dk^2}{k^2}\sqrt{1-\dfrac{4m_\chi^2}{k^2}}\left(1+\dfrac{2m_\chi^2}{k^2}\right)\\
    &\times \int dE_k\dfrac{1}{\sigma_{pN}}\dfrac{d\sigma_{pN}^{\rm PB}}{dE_k}\frac{\Theta(E_\chi-E_\chi^-)\Theta(E_\chi^+-E_\chi)}{E_\chi^+-E_\chi^-}\,,
\end{split}
\label{eq:PBMCPflux}
\end{equation}
where the second line corresponds to the decay rate of the virtual photon to a pair of MCPs, and the third line involves the differential cross section for proton bremsstrahlung with appropriate kinematical factors. The production rate is proportional to $d\sigma_{pN}^{\rm PB}/dE_k\propto \epsilon^2e^4$ (with one of the factors of $e^2$ inside $d\sigma^{\rm PB}_{pN}/dE_k$ in Eq.~\eqref{eq:PBMCPflux}), as is expected from bremsstrahlung. In more detail, the energy distribution of MCPs is described by a two-body decay boosted to the lab frame, where 
\begin{equation}
E_\chi^\pm=\gamma(E'_\chi\pm \beta |\pmb{p}'_\chi|)~,
\label{eq:Echipm}
\end{equation}
and the boost $\gamma=(1-\beta^2)^{-1/2}=E_k/\sqrt{k^2}$, with $E'_\chi$ and $|\pmb{p}'_\chi|$ the energy and momentum of MCPs in the rest frame of the virtual photon. We apply 
cuts to the virtual photon's four-momentum $k$ when computing the differential cross section $d\sigma^{\rm PB}_{pN}/dE_k$ such that only MCPs produced from the regime in which the FWW approximation is valid 
are included. Analogously to the rate of production of MCPs from meson decay, (given that $s'\simeq s$ and $\sigma_{pN}(s)$ has only a weak dependence on $s$) the proton-nitrogen cross section $\sigma_{p_N}$ cancels when Eq.~\eqref{eq:dsigmaPB} is substituted into Eq.~\eqref{eq:PBMCPflux}.

\subsection{Drell-Yan process}
The final MCP production channel that we consider is the Drell-Yan process illustrated in Figure~\ref{fig:feynman} right, i.e. the creation of a virtual $\gamma/Z$ through quark/anti-quark parton scattering that decays to a pair of MCPs. Drell-Yan is particularly important for MCPs with $m_\chi\ge 4.7$~GeV given the absence of intermediate mesons in this regime.  We obtain the corresponding differential cross section for the production of a pair of MCPs, $d\sigma^{\rm DY}(E_p)/dE_\chi$ by simulating the Drell-Yan process using \texttt{MadGraph~5}~\cite{alwall2014automated}, with the MCP added to the SM Lagrangian, and extracting the final state energy distribution statistics.

\begin{figure}[t!]
\centering
\includegraphics[width=\columnwidth]{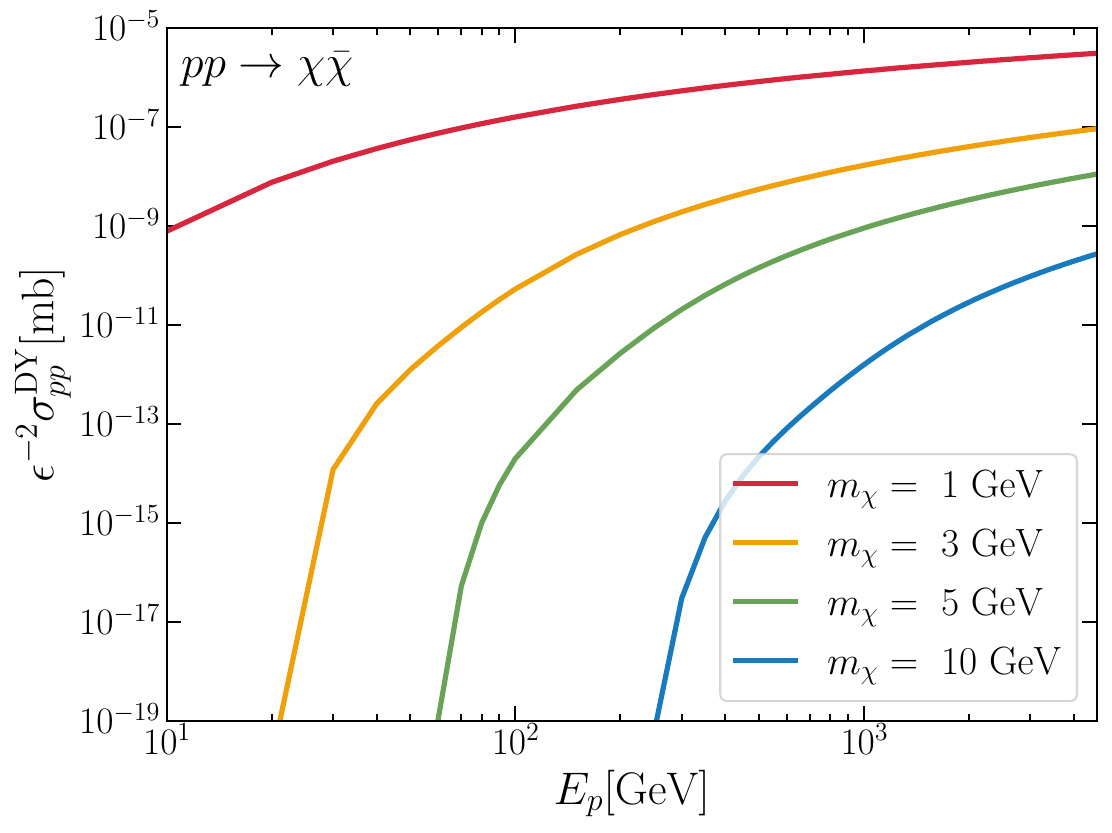}
\caption{The cross section $\sigma_{pp}^{\rm DY}$ for MCP production in the Drell-Yan process $pp\rightarrow\chi\bar{\chi}$ (illustrated in Figure~\ref{fig:feynman} right) as a function of the incident cosmic-ray proton energy $E_p$. Different colors correspond to different MCP masses $m_{\chi}$.}
\label{fig:drell-yan}
\end{figure}

The resulting cross section for MCP production by Drell-Yan is plotted in Figure~\ref{fig:drell-yan}. As an $s$-channel process, the cross section decreases fast for larger MCP masses. Meanwhile, for a fixed MCP mass, larger cosmic ray proton energy leads to a bigger cross section because this allows MCP production from a greater range of parton $x$ parameter space. 
The differential MCP flux from Drell-Yan is then straightforwardly obtained from the cosmic ray flux as 
\begin{equation} \label{eq:DY}
    \Phi^{\rm DY}_\chi\left(E_\chi\right)=2\int dE_p~\Phi_{p}\left(E_p\right)\frac{1}{\sigma_{pp}(E_p)}\dfrac{d\sigma^{\rm DY}_{pp}(E_p)}{dE_\chi}~,
\end{equation}
analogously to Eq.~\eqref{eq:dphidEm}.

\subsection{Millicharged particle flux}

\begin{figure}[t!]
\centering
\includegraphics[width=\columnwidth ]{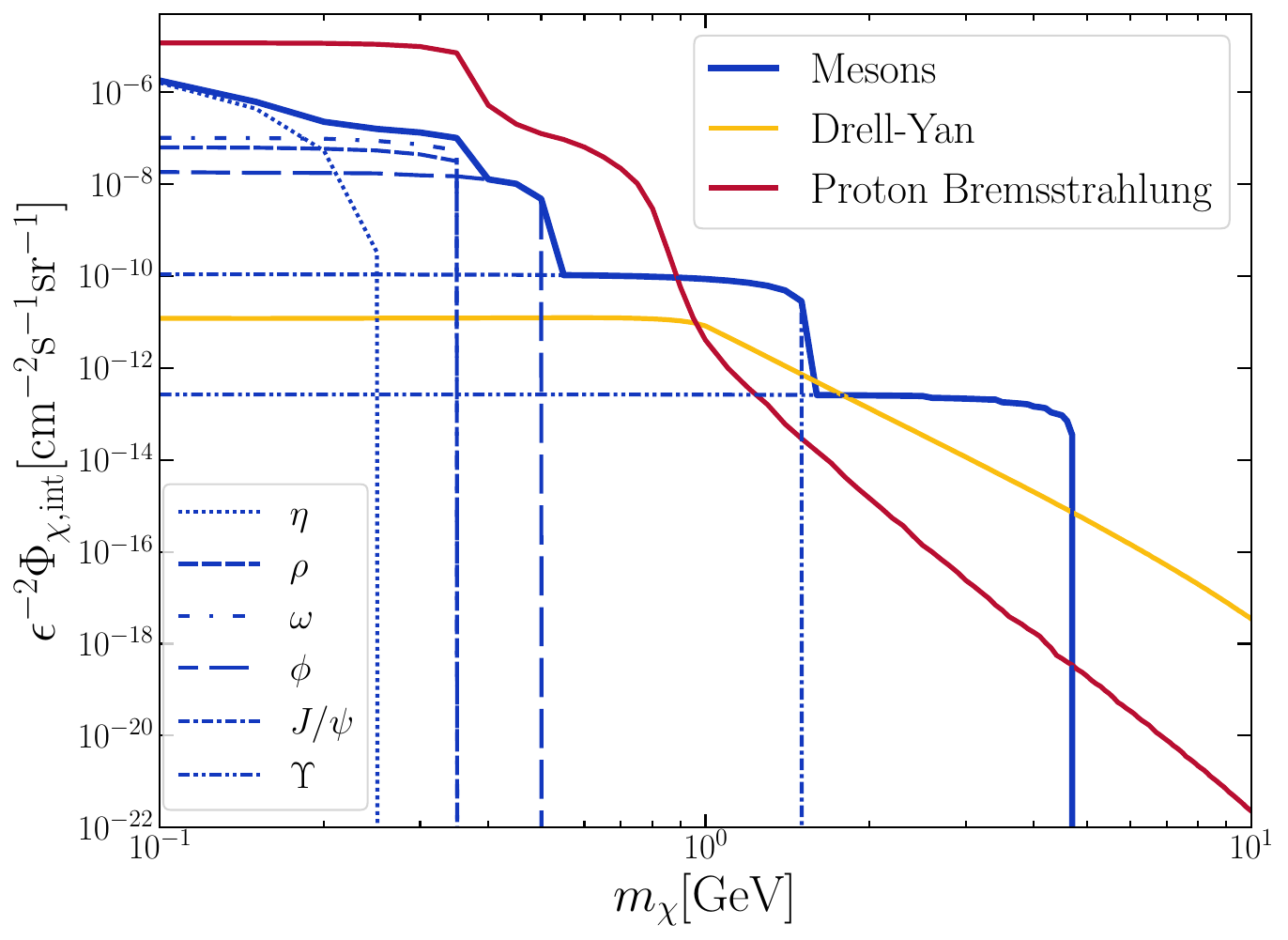}
\caption{The MCP flux from cosmic-rays interacting with the atmosphere as a function of MCP mass from different production processes, integrated over MCP energy. The blue solid line shows the MCP flux from the decay of mesons including the $\eta$, $\phi$, $\rho$, $\omega$, $J/\Psi$, and $\Upsilon$, with the individual contributions indicated by broken lines. The yellow and red lines correspond to the MCP flux from the Drell-Yan process and 
proton bremsstrahlung respectively.}
\label{fig:prod}
\end{figure}

In Figure~\ref{fig:prod} we plot the flux of MCPs integrated over all MCP energies, i.e. $\Phi_{\chi,{\rm int}}^{(i)}=\int dE_\chi~ \Phi_\chi^{(i)}(E_\chi)$, as a function of the MCP mass for the different processes labeled $(i)$. 
At low masses, $m_\chi\lesssim {\rm GeV}$, the MCP flux is mostly produced by proton bremsstrahlung. 
For $m_\chi \lesssim 0.3~{\rm GeV}$ the total flux from bremsstrahlung is approximately constant while for larger $m_\chi$, this production channel is suppressed by the proton electromagnetic form factor at large momentum transfer. 
Consequently, at intermediate masses, ${\rm GeV}\lesssim m_\chi\lesssim 4.7~{\rm GeV}$, meson decay takes over as the most important process. As the MCP mass is increased, the resulting MCP flux has clear steps at $m_\chi=m_\meson/2$, corresponding to the contributions from different mesons switching off. Finally, for $m_\chi > m_\Upsilon /2$, Drell-Yan dominates.

\section{Propagation and detection of millicharged particles}
\label{sec:detection}

\subsection{Attenutation}

Similarly to other charged particles, MCPs lose energy when propagating in the Earth through a variety of processes, including ionization, bremsstrahlung, electron-positron pair production and inelastic hadronic interactions. If a significant fraction of MCPs do not reach a detector, or if their energy spectrum is changed substantially, this can have a significant impact on the resulting sensitivity. Following Ref.~\cite{ArguellesDelgado:2021lek}, we model the relation between the energy of a MCP at the Earth's surface $E_\chi^s$ and when it reaches the detector $E_\chi^D$ by
\begin{equation} \label{eq:rel}
E_\chi^s\simeq (E_\chi^D+a_\chi/b_\chi)\exp(b_\chi X)-a_\chi/b_\chi~,
\end{equation}
where $X$ is the (water equivalent) distance traveled from the surface to the detector. In Eq.~\eqref{eq:rel}, $a_\chi$ and $b_\chi$, both approximately proportional to $\epsilon^2$, parameterize the rate of different energy loss processes such that $dE_\chi/dX \simeq -(a_\chi+b_\chi E_\chi)$. For MCPs with masses and energies in the range of interest, $0.1~{\rm GeV}\lesssim m_\chi \lesssim 10~{\rm GeV}$ and ${\rm GeV}\lesssim E_\chi \lesssim {\rm TeV}$, $a_\chi$ dominates. Being the result of ionizing scatterings, $a_\chi$ is approximately independent of the MCP mass (meanwhile $b_\chi$ might have a stronger dependence on $m_\chi$ \cite{hu2017dark} but has only a small effect). 
We therefore fix $a_\chi$ and $b_\chi$ to simply be the rescaled standard muon energy loss parameters for rock, i.e. $a_\chi / \epsilon^2 =  0.22 \ \mathrm{GeV\cdot mwe^{-1}} $ and $b_\chi/\epsilon^2 =  4.6\times10^{-4}\ \mathrm{mwe}^{-1} $ (see Appendix~\ref{app:attenuation} for details) \cite{koehne2013proposal}.

The flux of MCPs that reaches the detector $\Phi^D_\chi$ is  given by~\cite{gaisser2016cosmic}
\begin{equation} \label{eq:PhiD}
    \Phi^D_\chi(E^D_\chi,\Omega)\simeq e^{ b_\chi X}\Phi^s_\chi(E^s_\chi) ~,
\end{equation}
where $\Phi^s_\chi$ is the MCP at the surface of Earth and $e^{ b_\chi X}$ is the Jacobian of the relation between $E^D_\chi$ and $E^s_\chi$, which is close to $1$ for the $m_\chi$ and $\epsilon^2$ of interest. 
We assume that the MCP flux at the surface is isotropic, however, because the traveling distance depends on the arrival direction, $\Phi^D_\chi$ is a function of the solid angle. In practice, we treat a MCP as lost when $E_\chi\simeq m_\chi$, at which point Eq.~\eqref{eq:rel} inevitably breaks down.  
Not surprisingly, attenuation enhances the low energy spectrum at the cost of degrading the high energy flux.  For $\epsilon^2\gtrsim 10^{-2}$,  energy loss becomes efficient at kilometer length scales, which is the typical depth of neutrino experiments.

\subsection{Single scatter MCP signals}

MCPs scatter elastically with electrons and nuclei in a neutrino detector through photon-mediated interactions. Consequently, the corresponding matrix elements contain a factor of $Q^{-2}$, where $Q$ is the four-momentum transfer, and the scattering cross section is expected to be strongly enhanced in the low $Q^2$ regime and (for a fixed recoil energy) dominated by scattering with electrons. 
Indeed, a full calculation gives that the MCP-electron  differential scattering cross section is~\cite{Magill:2018tbb,ArguellesDelgado:2021lek}
\begin{equation} \label{eq:scatteringrate}
\begin{split}
  \frac{d\sigma_{\chi e}}{dE_r}= &\dfrac{1}{16\pi}\epsilon^2e^4 \times \\
  &\frac{(E_r^2+2E_\chi^2)m_e-\left((2E_\chi+m_e)m_e+m_\chi^2\right)E_r}{E_r^2m_e^2\left(E_\chi^2-m_\chi^2\right)}\,, 
\end{split}
\end{equation}
where $E_r$ is the electron recoil energy, which can take values between $0$ and $E_{r,\rm max}=(E_\chi^2-m_\chi^2)m_e/(m_\chi^2+2m_eE_\chi+m_e^2)$. Typically  MCPs from the atmosphere are relativistic with $E_\chi\gg m_\chi$ and also have $E_\chi\gg E_r$ so $d\sigma_{\chi e}/dE_r\propto 1/E_r^2$, i.e. the event rates are peaked at low recoil energy, consistent with the preceding discussion. As a result, detectors with lower thresholds are advantageous, at least until backgrounds (which are also largest at small $E_r$) become significant. The expected number of MCPs events in an energy bin $i$ covering the energy range $[E_{i,\min},E_{i,\max}]$ is given by
\begin{equation}
\begin{split}
    N_i\left(m_\chi,\, \epsilon\right)=N_eT&\int_{E_{i,\min}}^{E_{i,\max}}dE_r ~\epsilon_D(E_r)\\
    &\times\int dE_\chi d\Omega~\Phi_\chi^D\left(E_\chi,\Omega\right)\frac{d\sigma_{\chi e}}{dE_r}\,,
\end{split}
\end{equation}
where $\epsilon_D(E_r)$ is the detection efficiency, $T$ is the running time of the experiment and $N_e$ is the total number of electrons in the detector. We assume azimuthal symmetry so the MCP flux at the detector is only a function of zenith angle $\cos\theta$.

\subsection{Multiple scatter MCP signals}

The main low energy backgrounds to searches for MCPs, radiation and neutrinos, both have a greatly suppressed multi-hit signal rate (the former lead to spatially localised signals and the latter almost never have multiple interactions). As proposed in Ref.~\cite{ArguellesDelgado:2021lek}, the problem of large background rates at low recoil energies can therefore be ameliorated by considering multiple scatter signals, which allows much smaller detection threshold energies to be used.

The mean free path of a MCP in a detector between interactions that lead to an electron recoil energy of at least $E_{r,\min}$ 
is $\lambda=1/(n_e\sigma_{\chi e,\min})$, where $n_e$ is the electron number density and the scattering cross section~\cite{Magill:2018tbb}
\begin{equation}
    \sigma_{\chi e,\min}\simeq\frac{e^2\epsilon^2}{4 m_eE_{r,\min}}=2.6\times 10^{-25}\epsilon^2~{\rm cm}^2\frac{\rm MeV}{E_{r,\min}}\,.
\end{equation}
The energy lost by a MCP in a typical interaction is negligible. Therefore, 
the probability for a MCP to scatter once in a detector is obtained from the Poisson distribution: 
\begin{equation}
    P_1= 1-\exp\left(-\frac{L_{D}}{\lambda}\right)\,,
\end{equation}
where $L_D$ is the average length that MCP travels in the detector. Meanwhile, the probability of scattering at least twice is~\cite{ArguellesDelgado:2021lek}
\begin{equation}
    P_{n\ge 2}(E_{r,\min})=1-\exp\left(-\frac{L_{D}}{\lambda}\right)\left(1+\frac{L_{D}}{\lambda}\right)\,,
\end{equation}
and the expected number of single scatter, $N_{\rm single}$, and  multiple scatter, $N_{\rm multi}$, events are related by
\begin{equation}
N_{\rm multi}=N_{\rm single} \frac{P_{n\ge 2}(E_{r,\min})}{P_1(E_{r,\min})}~.
\label{eq:Nmulti}
\end{equation}

\subsection{Neutrino detectors}

We consider two experiments in this work: Super-Kamiokande (SuperK) and the  Jiangmen Underground Neutrino Observatory (JUNO). These have, respectively, the best current and near-future sensitivity to MCPs, offering a comprise between their detector mass and energy threshold (IceCube despite having a large volume has relatively large energy threshold).

SuperK is by far the largest neutrino detector that has an $\mathcal{O}$(MeV) energy threshold. The detector has a fiducial volume of about 22.5~kton filled with water and is shielded above by, on average, $1000~{\rm m}$ of rock~\cite{fukuda2003super}. We use the data taken from April 1996 until February 2014, in the four phases SK-I to SK-IV  with a total live-time of 4517 days~\cite{bays2012supernova,ArguellesDelgado:2021lek,Super-Kamiokande:2016yck}, covering electron recoil energies between 3.5~MeV and 88~MeV. 
To derive the single scatter constraints on MCPs from this data, we build the one-sided Poisson likelihood $\mathcal{L}$ following~\cite{ArguellesDelgado:2021lek}. Assuming a chi-squared distribution we require the test statistic
\begin{equation}
    \mathcal{TS}=-2\left[\dfrac{\mathcal{L}(m_\chi,\epsilon)}{\mathcal{L}(m_\chi,\epsilon=0)}\right]<2.71 ~,
\end{equation}
to obtain the 90\% upper limit on $\epsilon$. Multiple scatter signals cannot be searched for using SuperK data because there is not a suitable trigger system.

JUNO is located in a 700~m underground laboratory in Jiangmen, China and is expected to commence data taking in 2024. The JUNO main detector can be modeled as a sphere with an inner detector of 35.4~m, corresponding to  $L_D\simeq 23.6~{\rm m}$. The main detector contains 20~kton of liquid scintillator, enabling a much lower detection threshold than SuperK of about 100~keV. Moreover, an even lower-threshold electronic trigger system is being developed in order to reach thresholds of about 10~keV~\cite{JUNO:2021vlw} for multi-scatter events.

To obtain JUNO's projected single scatter sensitivity we use the sum of backgrounds in the energy range 10~MeV to 38~MeV (obtained from the right panel of Figure 5 in Ref.~\cite{JUNO:2021vlw}). Because this background estimate is still preliminary, rather than considering its energy dependence we simply use the total number of backgrounds with a 170 kton$\cdot$yr exposure. We set the threshold to $10$~MeV to avoid a large low-energy background dominated by solar neutrinos and the decay of cosmogenic isotopes~\cite{JUNO:2015zny} (extending the analysis window to lower energy does not improve the sensitivity to MCPs due to this high background rate). 

As mentioned, multiple scatter signals of MCPs allow for much lower thresholds. MCPs traverse the JUNO detector on times of order $10^{-7}$~s, which is shorter than the timescale over which the signal from the scintillator persists, roughly 200~ns. Within such a 200~ns time window, the background due to coincident cosmogenic radioactive decays is about 1 event per 10 years of data taking~\cite{ArguellesDelgado:2021lek}. At threshold energies below $100~{\rm keV}$ the photo-multiplier tube dark noise increases, however the background from this has not yet been analysed by the collaboration. We also note that the JUNO electronics have a timing accuracy of 200~ps~\cite{JUNO:2021vlw}, which could allow better timing resolution and further reduction of backgrounds (this would be important if backgrounds were larger than expected). 
Conservatively we obtain our projected sensitivity by requiring at least 10 multiple scatter events from MCPs with a 170 kton$\cdot$yr exposure regardless of threshold energy (changing this condition by order-one factors does not affect the sensitivity substantially).

\section{Results}
\label{sec:results}

The expected number of single scatter and multiple scatter events in a detector are obtained by combining our preceding results for the MCP flux at the Earth's surface, the attenuation (accounting for the varying MCP incident angle) with the scattering cross section in Eq.~\eqref{eq:scatteringrate}. 

\begin{figure}[!t]
\centering
\includegraphics[width=\columnwidth]{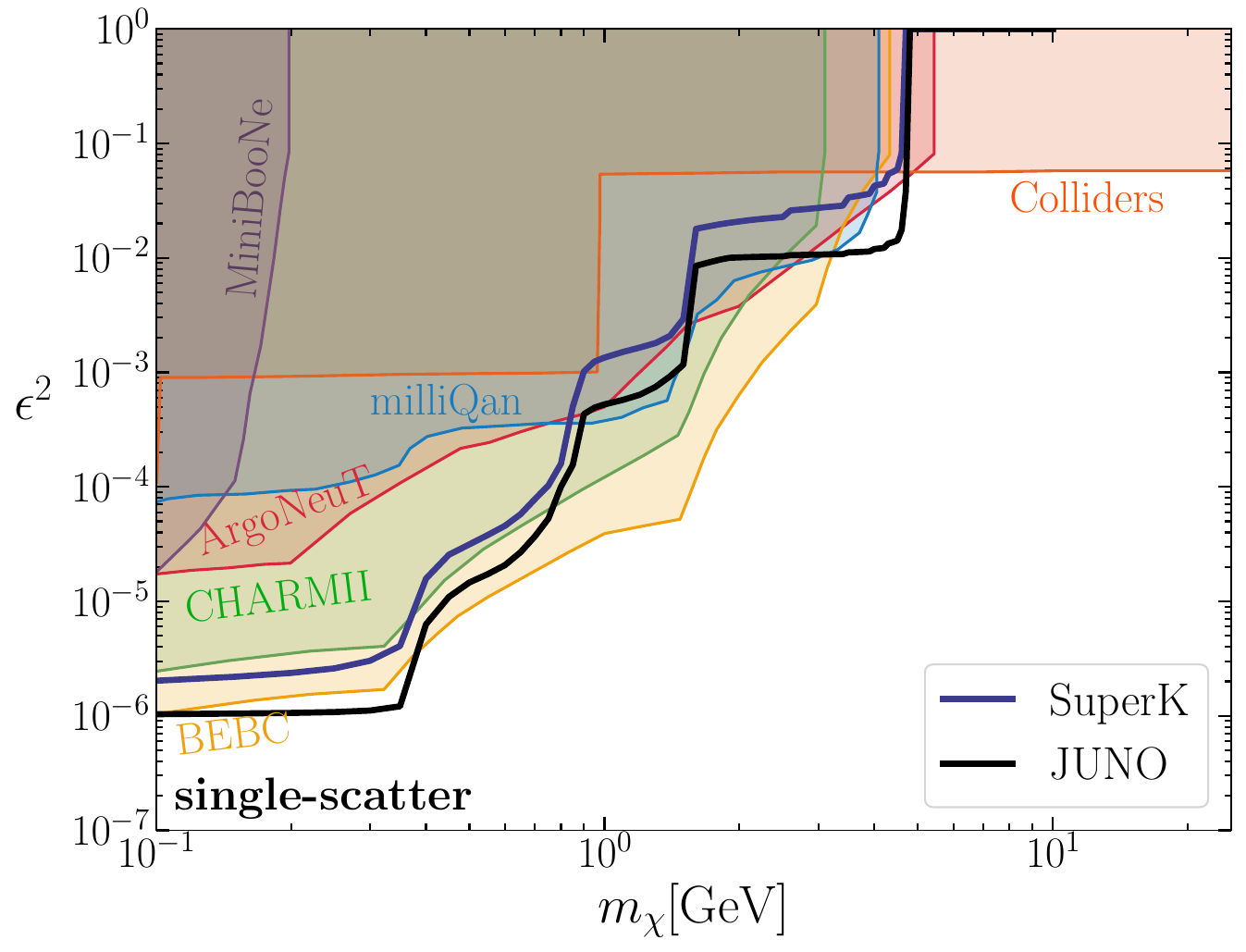}
\caption{Constraints on MCPs from single scatter events at SuperK (using the SK-I to SK-IV data), and the projected sensitivity of JUNO assuming an exposure of 170 kton$\cdot$yr. Overlaid are the same existing constraints as in Figure~\ref{fig:results_multi}.}
\label{fig:results_single}
\end{figure}

The resulting constraints from SuperK and the projected sensitivity of JUNO due to single scatter signals are shown in Figure~\ref{fig:results_single}. As expected from Figure~\ref{fig:prod}, as $m_\chi$ is increased the sensitivity is, in turn, dominated by production from proton bremsstrahlung, meson decay and finally Drell-Yan. 
For MCP masses such that bremsstrahlung gives the largest contribution,  the constraint on $\epsilon^2$ that we obtain from our conservative approach is comparable to that using the new method of Ref.~\cite{Du:2022hms}, and is approximately a factor of 2 stronger than obtained using the FWW approximation in that reference (this might be due to differences in the computation of the proton bremsstrahlung MCP flux or the likelihood). 
The dominance of bremsstrahlung for small MCP masses is not too surprising given that Ref.~\cite{deNiverville:2016rqh} found that this could also occur for light dark matter production at fixed target experiments in similar kinematic ranges. 
Although the current SuperK data does not rule out any new parameter space, JUNO has potential sensitivity to a small region of otherwise unconstrained parameter space for $2.8~{\rm GeV}\leq m_\chi \leq 4.7~{\rm GeV}$.

Meanwhile, the projected sensitivity to MCP multiple scatter events at JUNO is shown in Figure~\ref{fig:results_multi}. 
We plot results for different values of the threshold $E_{\rm th}$ to illustrate its importance; the sensitivity is dramatically enhanced when the threshold is reduced from $1~{\rm MeV}$ to $10~{\rm keV}$. For $E_{\rm th}=1~{\rm MeV}$, attenuation (which shifts the MCP flux to smaller energies) destroys the sensitivity for $m_\chi>4.7~{\rm GeV}$ while for lower $E_{\rm th}$ sensitivity is maintained until progressively larger $m_\chi$. 
In the optimistic case of $E_{\rm th}=10~{\rm keV}$, JUNO's sensitivity surpasses current constraints on $\epsilon^2$ by up to three orders of magnitude. 

We present a comparison of our results for single and multiple scatter searches with those in previous work in Appendix~\ref{app:compare}.

\section{Conclusions}
\label{sec:discussions}

In this paper we have revisited the current constraints and future sensitivity of neutrino detectors to MCPs from the atmosphere. 
While current and future single scatter searches yield sensitivities that are comparable to existing limits, multiple scatter searches at JUNO could far surpass current bounds on the MCP charge. The projected sensitivity is particularly good relative to existing constraints in the range of MCP masses, $m_\chi \gtrsim 2~{\rm GeV}$, in which production by $\Upsilon$ decay and Drell-Yan is relevant (which we have included for the first time). Our results therefore further motivate dedicated multiple scatter searches at JUNO and other neutrino detectors.

There are various possible future improvements that would allow neutrino detectors to reach better sensitivity to MCPs. On the experimental side, a lower detection threshold and faster detector response could increase the MCP signals and reduce the coincident background. Better background modelling and analysis would help with background reduction. Furthermore, dedicated multiple scatter simulations are required to obtain a precise determination of the signal and background rates. On the theoretical side, the calculation of the MCP production rate from bremsstrahlung using the FWW and splitting kernel approximation requires relativistic and collinear conditions to be satisfied, such that the MCP flux is underestimated. It would be useful if more complete methods could be developed to account for the full kinematic range given that bremsstrahlung dominates for small MCP masses.

Finally, we stress that searches for MCPs from the atmosphere at neutrino detectors are by no means the only route to improved sensitivity to MCPs with masses in the range $0.1~{\rm GeV}\lesssim m_\chi \lesssim 10~{\rm GeV}$. In this regard we highlight Refs.~\cite{Haas:2014dda,Kelly:2018brz,Choi:2020mbk,Foroughi-Abari:2020qar} which propose dedicated detectors searching for MCPs in the forward region of particle colliders. These could lead to spectacular sensitivity, potentially as good as $\epsilon^2 \simeq 10^{-7}$ for $m_\chi\lesssim 40~{\rm GeV}$, which far exceeds current constraints. Nevertheless, since JUNO is being developed and run for other purposes, it remains valuable to exploit its full physics potential by using the resulting data to search for MCPs in addition to its core neutrino programme.

\section*{Acknowledgements}
We thank Christopher Cappiello, Mingxuan Du, Juri Fiaschi, Qinrui Liu, Zuowei Liu and Aaron Vincent for useful discussions. NS also acknowledges Yufeng Li and Liangjian Wen for correspondence on the JUNO experiment. 
HW is supported by the Arthur B. McDonald Canadian Astroparticle Physics Research Institute, the Canada Foundation for Innovation and the Queen’s Centre for Advanced Computing. 
EH acknowledges the UK Science and Technology Facilities Council for support through the Quantum Sensors for the Hidden Sector collaboration under the grant ST/T006145/1 and UK Research and Innovation Future Leader Fellowship MR/V024566/1. 
NS is supported by the National Natural Science Foundation of China (NSFC) No. 12347105, No. 12475110 and No. 12047503.

\appendix
\section{Details of MCPs Production from Meson Decay}
\label{app:mesondecay}

\subsection{Cross sections for meson production} 

For all the relevant mesons except for the $\Upsilon$, we use the production cross sections provided in Ref.~\cite{Plestid:2020kdm}, which are obtained by fitting collider data. 

\begin{figure}[!t]
    \centering
    \includegraphics[width=\columnwidth]{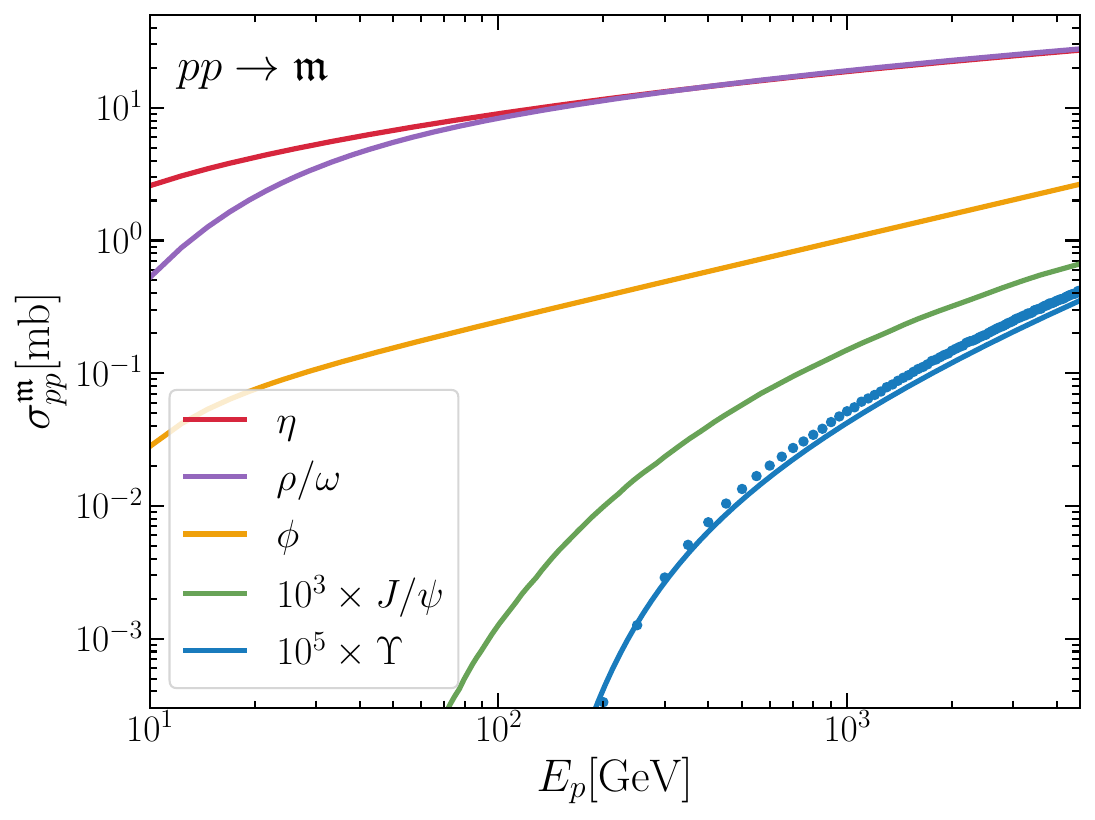}
    \caption{Cross sections for meson production in \textit{pp} collisions $\sigma_{pp}^\meson$  where $E_p$ is the incident proton energy. For the $\Upsilon$ we show numerical data (dots) and our fit of Eq.~\eqref{eq:Upfit}.}
    \label{fig:mesoncrosssections}
\end{figure}

As mentioned in the main text, we calculate the $\Upsilon$ production cross section in $pp$ collisions using \texttt{Pythia~8.3} (in particular, we turn off $\Upsilon$ decays and extract the differential cross section for $\Upsilon$ production in $pp$ collisions from the appearance of final states that include an $\Upsilon$). We find that the resulting cross section is well fit by
\begin{equation}
\begin{split} \label{eq:Upfit}
    \ln\left(\dfrac{\sigma_{pp}^{\Upsilon}}{\rm mb}\right)=&-102.6+78.6(\ln\gamma_{\rm cm})^{0.1}\\
    &\times\left(1+\dfrac{0.76}{(\gamma_{\rm cm}-5.04)^2}\right)^{-1} ~,
\end{split}
\end{equation}
where the boost of the $pp$ center of mass $\gamma_{\rm cm}=s/(2m_p)=\sqrt{2E_pm_p}/(2m_p)$. 
We plot the total cross sections for meson production in proton-proton collisions as a function of the incoming proton energy $E_p$ in Figure~\ref{fig:mesoncrosssections}.

In order to calculate the spectrum of MCPs produced by meson decay (by evaluating Eq.~\eqref{eq:dphidEm}), we require the differential meson production cross section $d\sigma_{pp}^\meson/dE_\meson$. Rather than $E_\meson$, the differential cross section is typically expressed in terms of $x_F$, which is the Feynman-$x$ parameter defined as $x_F \equiv p'_{\parallel}/p_{\rm max}$ where $p'_{\parallel}$ ($p_{\rm max}$) is the (maximum possible) meson longitudinal momentum in the $pp$ collision center of mass frame. For subsequent use, we note that $p_{\max}=\sqrt{s}\left(1-m_\meson^2/s\right)/2$ and that $x_F$ is related to the meson boost $\gamma_\meson$ by 
\begin{equation}
    \gamma_\meson\simeq \dfrac{\gamma_{\rm cm}p_{\max}}{m_\meson}\left(\sqrt{x_F^2+\dfrac{m_\meson^2}{p_{\max}^2}}+\beta_{\rm cm}x_F\right)\,,
    \label{eq:gammam}
\end{equation}
with $\beta_{\rm cm}$ the velocity of the center of mass of the system.

Still following Ref.~\cite{Plestid:2020kdm} (which should be consulted for plots of the original data and a discussion of the fitting approach and the uncertainties), the differential cross section for $\eta$ meson production in proton proton collisions  $d\sigma_{pp}^\eta/dx_F$ is reasonably well fit by
\begin{equation} \label{eq:diff}
   \frac{d\sigma_{pp}^\eta}{dx_F}=\sigma_{pp}^\eta\times\frac{c_\eta/2}{1-\exp(-c_\eta)}\exp\left(-c_\eta|x_F|\right)\,,
\end{equation}
where $c_\eta=9.5+(\gamma_{\rm cm}-14.6)/2$. The differential cross sections for production of the vector mesons $\rho$, $\omega$ and $\phi$ can be well fit by the same functional form as Eq.~\eqref{eq:diff}, with $c_\eta$ replaced by 
$c_V=7.7+0.44\left(\gamma_{\rm cm}-14.6\right)$. 
The differential cross section of $J/\psi$ can be approximated as~\cite{vogt1999j}:
\begin{equation}
   \frac{d\sigma_{pp}^{J/\psi}}{dx_F}=\sigma_{pp}^{J/\psi}\times\frac{c_{J/\psi}+1}{2}\left(1-|x_F|\right)^{c_{J/\psi}}~,
\end{equation}
with $c_{J/\psi}\left(\gamma_{\rm cm}\right)=2+0.2(\gamma_{\rm cm}-5)$. 

\begin{figure*}[!htb]
\centering
\includegraphics[width=0.99\columnwidth]{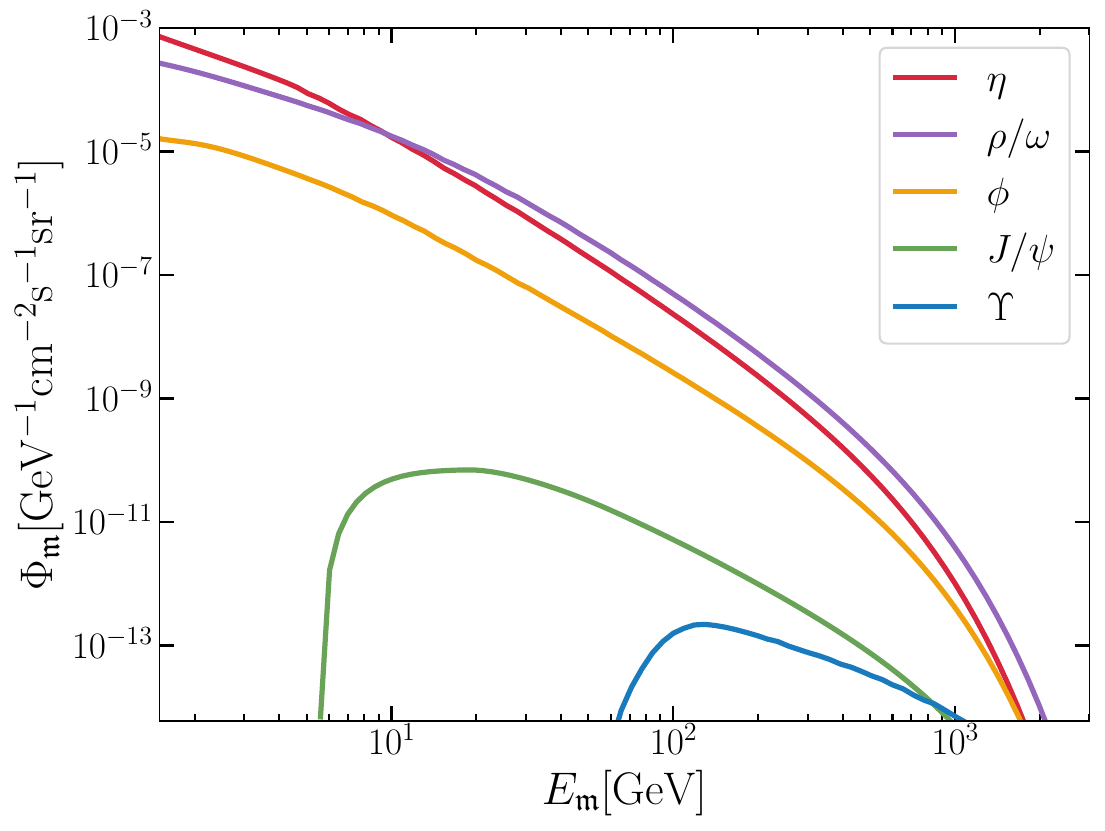} \quad
\includegraphics[width=0.99\columnwidth]{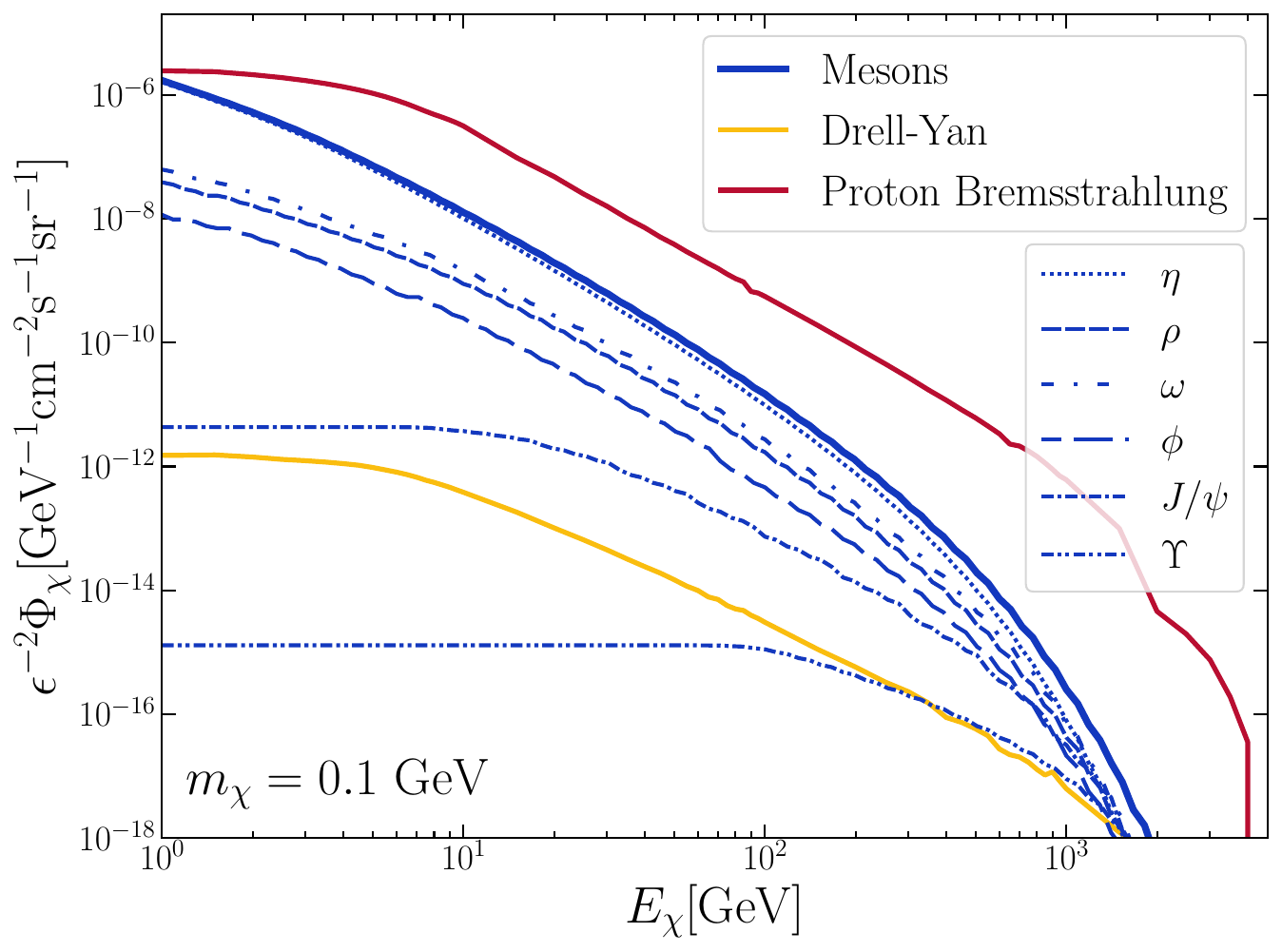}
\caption{{\it  Left:} The differential flux of mesons $\Phi_\meson$ produced by cosmic rays collisions with the atmosphere as a function of the meson energy $E_{\rm m}$. The fluxes of $\rho$ and $\omega$ are very similar so we show both as a single line.   {\it Right:} The differential flux of MCPs $\Phi_\chi$ from the atmosphere produced by different processes (for production from meson decay, the contributions from different mesons are also indicated) as a function of the MCP energy $E_\chi$ with the MCP mass $m_\chi=0.1~{\rm GeV}$ fixed. }
\label{fig:MCPproduction}
\end{figure*}

To obtain the differential cross section in terms of $E_\meson$, we invert Eq.~\eqref{eq:gammam}, which leads to two $x_F$ branches
\begin{equation}
   x_F^{\pm}=-\gamma_{\rm cm}\gamma_\meson\left(\beta_{\rm cm}\pm\beta_\meson\right)\frac{m_\meson}{p_{\max}}\,.
   \label{eq:xFpm}
\end{equation}
The integral in Eq.~\eqref{eq:dphidEm} is then performed by changing variables from $E_\meson$ to $x_F$ and summing the two branches of $x_F$ with the Jacobian $dx_F^\pm/dE_\meson$ included. 
By definition, $x_F$ ranges from $-1$ to $1$. However, this range is not fully kinematically accessible for both $x_F^+$ and $x_F^-$. When the meson is produced at rest, i.e. $\gamma_\meson=1$, we find from Eq.~\eqref{eq:gammam} and~\eqref{eq:xFpm} the solution $x_F=x_{F0}\equiv -\beta_{\rm cm}\gamma_{\rm cm}m_\meson/p_{\max}$ (it can be verified that the $x_F>0$ solution from Eq.~\eqref{eq:xFpm} is spurious in this case). As $x_F^+$ decreases from $x_{F0}$ and $x_F^-$ increases from $x_{F0}$ the meson boost in the lab frame $\gamma_\meson$ grows. Therefore, the integral is restricted to the ranges $-1 \le x_F^+\le x_{F0}$ and $x_{F0} \le x_F^-\le 1$.

\subsection{Meson decays to MCPs}

The branching ratio of MCPs from two-body vector meson decay is similar to the dimuon decay channel, but with a different mass and coupling. Accounting for the coupling and phase space differences, we have \cite{Plestid:2020kdm,ArguellesDelgado:2021lek} 
\begin{equation}    
\dfrac{{\rm Br}\left(\meson\rightarrow\chi\bar\chi\right)}{{\rm Br}\left(\meson\rightarrow\mu^+\mu^-\right)}=\epsilon^2\dfrac{m_\meson^2+2m_\chi^2}{m_\meson^2+2m_\mu^2}\sqrt{\frac{m_\meson^2-4m_\chi^2}{m_\meson^2-4m_\mu^2}}\,,
\end{equation}
The branching ratios for $\rho$, $\omega$, $\phi$ $J/\psi$ and $\Upsilon$s decaying to muon pairs are $4.55\times 10^{-5}$, $7.40\times 10^{-5}$, $2.87\times 10^{-4}$, $5.96\times 10^{-2}$, $2.48\times 10^{-2}$ respectively \cite{Plestid:2020kdm,particle2022review}. In the rest frame of the meson, the energy of the produced MCPs is simply $m_\meson/2$. After boosting, the distribution of final MCP energy in the lab frame is flat between the maximum and minimum possible MCP energies $E_\chi^\pm$ \cite{Plestid:2020kdm}, i.e.
\begin{equation} \label{eq:trans}
     P\left(E_\chi| E_\meson\right) \equiv \frac{1}{\Gamma_\meson}\frac{d\Gamma_\meson}{dE_\chi}=
           \frac{1}{E_\chi^+-E_\chi^-} ~,
\end{equation} 
with $E_\chi^\pm$ defined analogously to in Eq.~\eqref{eq:Echipm} with $\gamma$ replaced by the meson boost $\gamma_\meson$. The MCP energy in the meson rest frame $E'_\chi=m_\meson/2$ for $m_\chi\le m_\meson/2$.

In the case of pseudoscalar mesons the relevant process is $\meson\rightarrow \gamma\chi\chi$. The only pseudoscalar meson we consider is the $\eta$, and the branching ratio of this to MCPs can be related to that of $\eta\rightarrow\gamma\gamma$ by~\cite{Plestid:2020kdm,ArguellesDelgado:2021lek}
\begin{equation}
{\rm{Br}}\left(\eta\rightarrow\gamma\chi\chi\right)= \frac{1}{2\pi}e^2\epsilon^2 {\rm{Br}}(\eta\rightarrow\gamma\gamma)I^{(3)}\left(\frac{m_\chi^2}{m_\eta^2}\right)\,,
\end{equation}
where ${\rm Br}\left(\eta\rightarrow\gamma\gamma\right)=3.941\times 10^{-1}$ and the dimensionless function $I^{(3)}(x)$ is given by
\begin{equation}
    I^{(3)}(x)=\frac{2}{3\pi}\int_{4x}^1 dz ~\sqrt{1-\frac{4x}{z}}\frac{(1-z)^3}{z^2}\left(2x+z\right)\,.
\end{equation}

The energy distribution of the MCPs from such decays can be obtained following Ref.~\cite{Plestid:2020kdm,ArguellesDelgado:2021lek}. In particular, we use the analytical formalism implemented in the \texttt{HeavenlyMCP} code~\cite{ArguellesDelgado:2021lek}, which is obtained by boosting the distribution in the meson rest frame to the lab frame. Defining $z\equiv E_\chi/\gamma_\eta$, the energy distribution
\begin{equation}
   P(E_\chi|E_\meson)\equiv \frac{1}{\Gamma_\eta\gamma_\eta}\frac{d\Gamma_\eta}{dz}=\frac{m_\eta-z}{72z^3F_1(m_\chi)}F_2\left(z,m_\chi\right) ~,
\end{equation}
where 
\begin{align}
\begin{split}
F_1=&\frac{1}{24m_\eta}[-M_\chi^8+8M_\chi^6m_\eta^2-24M_\chi^4m_\eta^4\ln\frac{M_\chi}{m_\eta}\\
&-8M_\chi^2m_\eta^6+m_\eta^8]\,,
\end{split}\\
\begin{split}F_2=&M_\chi^6(4m_\eta^2-5m_\eta z-5z^2)\\
   &-9M_\chi^4m_\eta^2z(m_\eta-3z)\\
   &+9M_\chi^2m_\eta^2z^3(z-3m_\eta)\\
   &+m_\eta^3z^3(5m_\eta^2+5m_\eta z-4z^2)\,,
   \end{split}
\end{align}
with $M_\chi=2m_\chi$, and $z$ in the range
\begin{equation}
   E_{\max}-\sqrt{E_{\max}^2-M_\chi^2}\le z \le E_{\max}+\sqrt{E_{\max}^2-M_\chi^2} ~,
\end{equation}
where $E_{\max}\equiv(M_\chi^2+m_\eta^2)/(2m_\eta)$.

\section{MCP production from bremsstrahlung}
\label{app:protonbrem}

\begin{figure*}[!t]
\centering
\includegraphics[width=1\columnwidth]{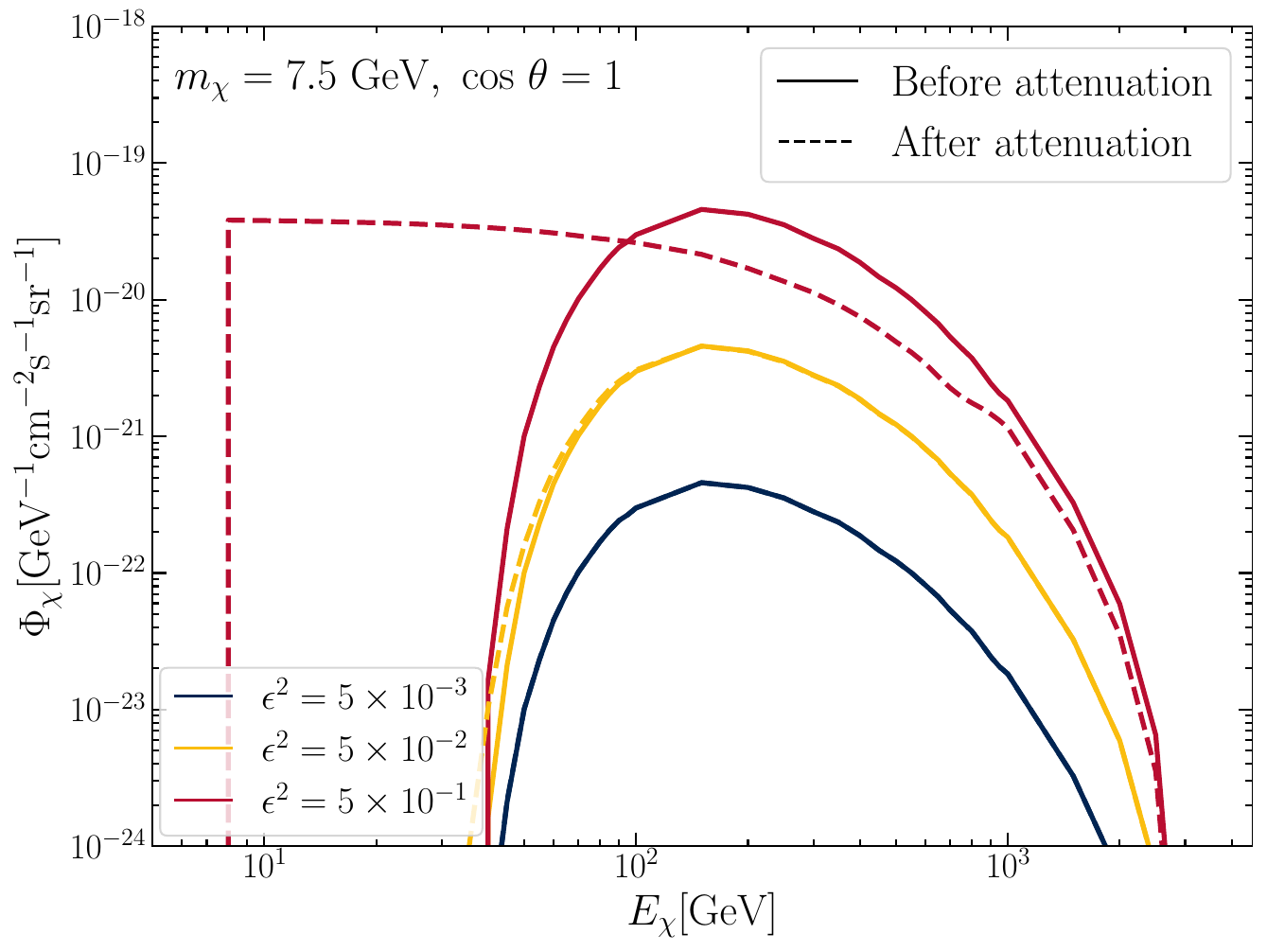}
\includegraphics[width=1\columnwidth]{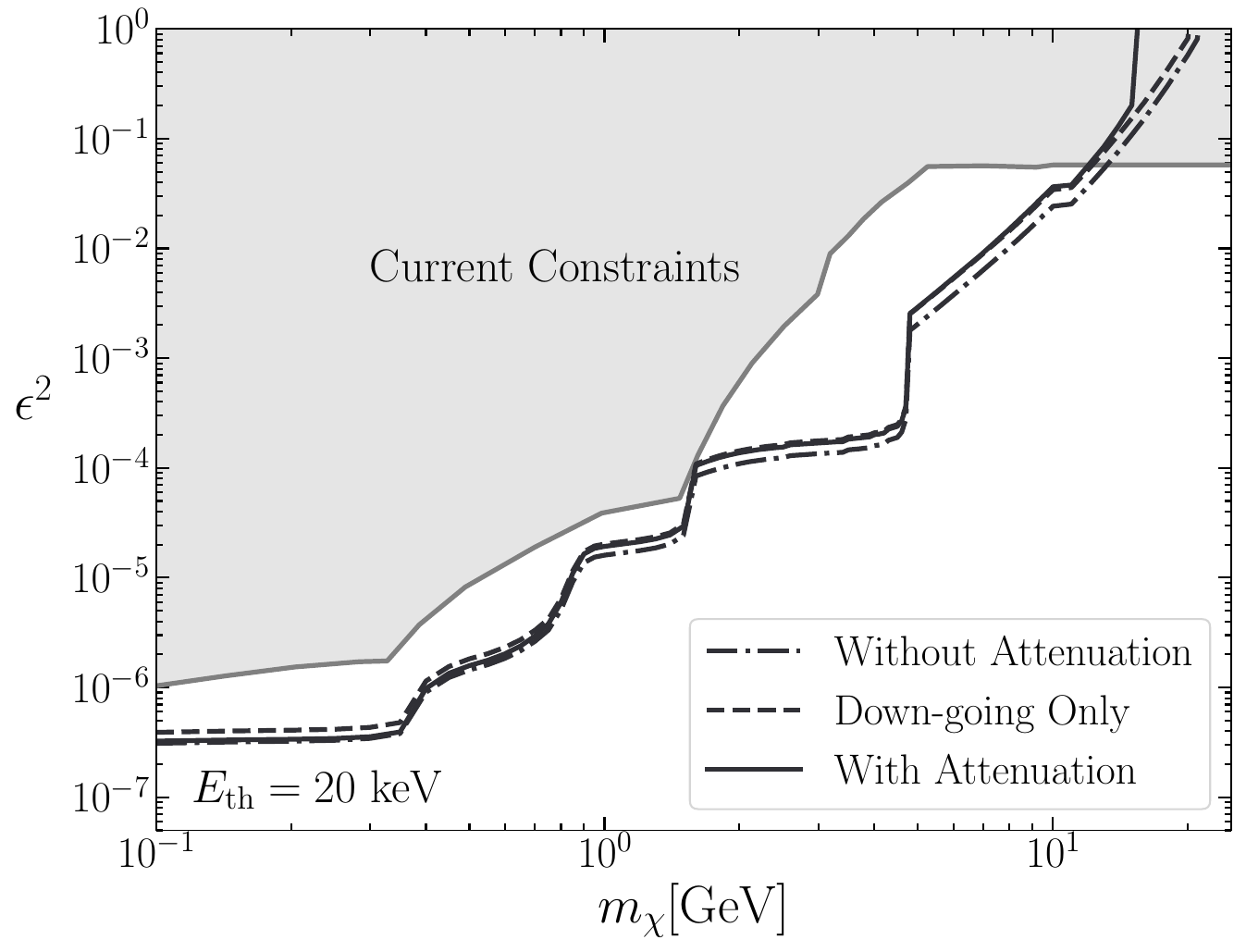}
\caption{{\it Left:} The impact of attenuation by the Earth on the MCP flux $\Phi_\chi$ as a function of the MCP energy $E_\chi$ for a detector at the depth of JUNO,  with $m_\chi=7.5$~GeV fixed, for different values of $\epsilon^2$ at the zenith angle $\cos\theta=1$ (i.e. for MCPs from directly above). Given the relatively large MCP mass, the MCPs are only produced by the Drell-Yan process and have a minimum boost at about 40~GeV. For $\epsilon^2=5\times 10^{-2}$ and $5\times 10^{-1}$ attenuation has a significant effect on $\Phi_\chi$, shifting the flux to lower energy. The attenuated flux is truncated below the MCP mass, because Eq.~\eqref{eq:dedx} stops being accurate once $E_\chi\simeq m_\chi$, and we regard the MCP as being stopped at this point. 
{\it Right:} The sensitivity of JUNO to MCPs using a multiple scatter analysis with threshold energy $E_{\rm th}=20~{\rm keV}$ for different treatments of the attenuation. The solid line corresponds to the full treatment (used to obtain the results in the main text), the dashed line considers only the down-going MCP flux assumed to enter the detector unimpeded, while the dash-dotted line shows the sensitivity without including any attenuation.}
\label{fig:Attenuation}
\end{figure*}

In FWW approximation, the splitting probability of a proton of four-momentum $p$ is~\cite{Blumlein:2013cua,deNiverville:2016rqh,Foroughi-Abari:2021zbm,Gorbunov:2023jnx,Du:2022hms}
\begin{equation} 
    d^2\mathcal{P}_{p\rightarrow \gamma^*p^\prime}=\omega(z,p_T^2)|F_V(k^2)|^2dzdp_T^2\,,
    \label{eq:dPsplitting}
\end{equation}
where $k^\mu=(E_k,\pmb{k})$ is the emitted virtual photon four-momentum. In Eq.~\eqref{eq:dPsplitting} we use the photon's transverse momentum $p_T$ and $z$ as our kinematic variables; these are related to $p$, $k$ and $\theta_k$ (the angle between $\pmb{p}$ and $\pmb{k}$) by $p_T \equiv |\pmb{k}|\sin\theta_k$ and $z\equiv \cos\theta_k|\pmb{k}|/|\pmb{p}|$. The function $\omega$ is given by
\begin{equation}
\begin{split}
    \omega=\dfrac{e^2}{8\pi^2 H}& \left\{ \dfrac{1+z'^2}{z}-2zz'\left(\dfrac{2m_p^2+k^2}{H}-\dfrac{2m_p^4z^2}{H^2}\right)\right.\\
    &\left.+2zz'\left(z+z'^2\right)\dfrac{m_p^2k^2}{H^2}+2zz'^2\dfrac{k^4}{H^2}\right\}\,,
\end{split}
\end{equation}
where $z'=1-z$ and $H=p_T^2+z'k^2+z^2m_p^2$. As in Refs.~\cite{Du:2022hms,Foroughi-Abari:2021zbm}, in Eq.~\eqref{eq:dPsplitting} we include $F_V$, which is a hadronic form factor accounting for enhancement arising from mixing between the virtual photon and vector mesons. In particular, we take the phenomenological fit
\begin{equation}
    F_V(k^2)=\sum\limits_V \dfrac{f_Vm_V^2}{m_V^2-k^2-im_V\Gamma_V}\,. 
\end{equation}
where $V=\{\rho,\,\rho',\,\rho'',\,\omega,\,\omega',\,\omega''\}$ (with $'$ representing higher resonance states). For $\rho$ mesons the mass parameters $m_\rho/{\rm GeV}=\{0.775,\,1.464,\,1.570\}$, the decay widths $\Gamma_\rho/{\rm GeV}=\{0.1474,\,0.400,\,0.144\}$, and a fit to data gives $f_\rho=\{0.616,\,0.223,\,-0.339\}$. For $\omega$ mesons $m_\omega/{\rm GeV}=\{0.783,\,1.410,\,1.670\}$, $\Gamma_\omega/{\rm GeV}=\{0.0086,\,0.290,\,0.315\}$, and $f_\omega=\{1.011,\,-0.881,\,0.369\}$~\cite{Du:2022hms,particle2022review}.

Because the FWW and splitting kernel approximation only work when the initial and final state particles are relativistic and collinear, we consider only MCPs produced from scatterings in which the following conditions are met~\cite{Du:2022hms,deNiverville:2016rqh}: {\it i)} $p_T<0.1 E_k$, {\it ii)} $p_T<1$~GeV, {\it iii)} $|q^2_{\min}|\simeq\left(p_T^2+z'k^2+z^2m_p^2\right)^2\left(4E_p^2z^2z'^2\right)<\Lambda_{\rm QCD}^2$, with the QCD scale $\Lambda_{\rm QCD}\simeq 0.25$~GeV, and iv) $E_k>5m_p,\ E_p-E_k>5m_p$, $E_k>5\sqrt{k^2},\ E_p-E_k>5\sqrt{k^2}$. These conditions fix the allowed ranges of $\cos\theta_k$, $E_k$ and $k^2$.

\section{Attenuation of MCPs in the Earth}
\label{app:attenuation}

The average rate of energy loss as a MCP of mass $m_\chi$ travels through the Earth can be modeled by~\cite{hu2017dark,ArguellesDelgado:2021lek}
\begin{equation}
\begin{split} \label{eq:dedx}
    -\frac{dE_\chi}{dX}=&\epsilon^2(a_{\rm ion}+\epsilon^2b_{\rm el-brem}E_\chi+b_{\rm inel-brem}E_\chi\\
    &+b_{\rm pair}E_\chi+b_{\rm photo-had}E_\chi)\simeq a_\chi+b_\chi E_\chi~,
\end{split}
\end{equation}
where $X$ is the distance traveled, $a_x$ and $b_x$ represent energy loss processes (in particular, ionization, elastic and inelastic bremsstrahlung, electron-positron pair production, and photohadronization), which are summarized in $a_\chi$ and $b_\chi$. The values of $a_\chi$ and $b_\chi$ can be estimated from the corresponding parameters for muon energy loss~\cite{hu2017dark}. The ionization energy loss described by the Bethe formula is roughly independent of the particle mass but related to its charge. As mentioned in the main text, $b_\mu$ may have a stronger dependence on the MCP mass \cite{hu2017dark} but the corresponding energy loss for a MCP is negligible compared to that from $a_\chi$ for MCP masses in the range $0.1$ to $10$~GeV. We therefore simply take
\begin{equation}
\frac{a_\chi}{a_\mu}\simeq   \frac{b_\chi}{b_\mu} \simeq \epsilon^2\, .
\end{equation}
Given the dominance of $a_\chi$, the value of $m_\chi$ does not affect the energy loss significantly. The distance a MCP travels before reaching a detector is given by 
\begin{equation}
\label{eq:distance}
    X=\sqrt{\left(R_{\rm E}-d\right)^2\cos^2\theta+d\left(2R_{\rm E}-d\right)}-\left(R_{\rm E}-d\right)\cos\theta~,
\end{equation}
where $d$ is the depth of the detector, $R_{\rm E}$ is radius of the Earth and $\theta$ is the MCP incoming zenith angle. The flux of MCPs reaching the detector after attenuation is given by Eq.~\eqref{eq:PhiD} in the main text (see also~\cite{gaisser2016cosmic}).

The effect of attenuation by the Earth is illustrated in the left panel of Figure~\ref{fig:Attenuation}. As $\epsilon^2$ increases, the MCP energy is degraded more and more and the flux is shifted to progressively lower energies. For $\epsilon^2\gtrsim 10^{-2}$,  attenuation becomes significant even for MCPs arriving from overhead.

The impact of attenuation on the sensitivity of neutrino detectors is shown in the right panel of Figure~\ref{fig:Attenuation}. In particular, we compare the sensitivity of JUNO in three scenarios. In the first case, we assume that the MCP flux arrives at the detector without any attenuation. In the second case, we neglect attenuation but reduce the total MCP flux by a factor of 2, which is equivalent to assuming that the down-going MCP flux reaches the detector unaffected and all the up-going flux is lost. In the last case, we use the full attenuation treatment described in this section and used the main text. For $\epsilon^2 \lesssim 10^{-6}$ the results with attenuation included and ignoring attenuation match closely (and the down-going approximation reduces the flux too much). 
For $10^{-6}\lesssim\epsilon^2\lesssim10^{-1}$, the down-going approximation matches the full results well. At larger $\epsilon^2$ the down-going flux is also attenuated and the full sensitivity is worse than suggested by the down-going approximation.

\section{Comparison with previous work}
\label{app:compare}

\begin{figure*}
    \centering
    \includegraphics[width=0.49\linewidth]{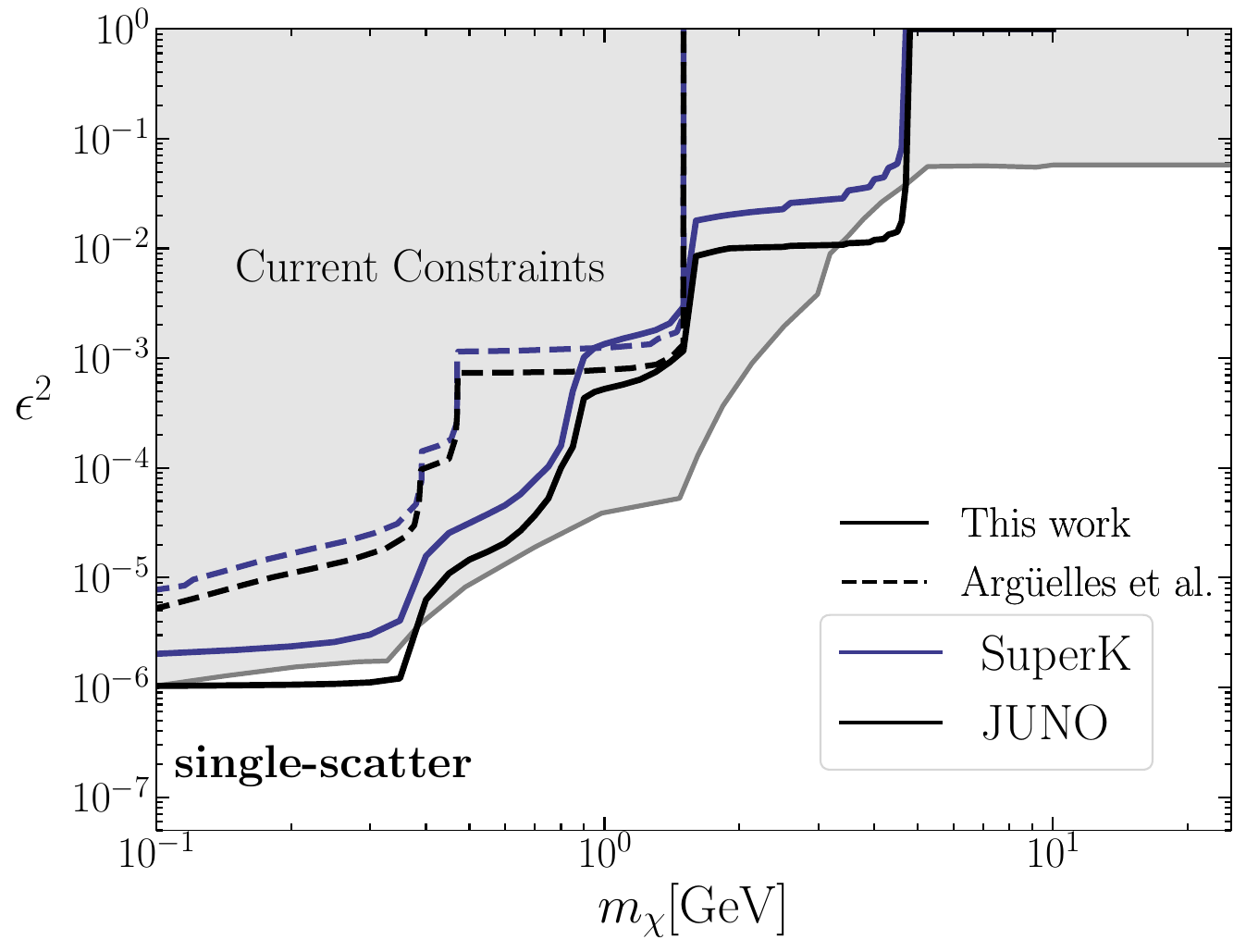}
    \includegraphics[width=0.49\linewidth]{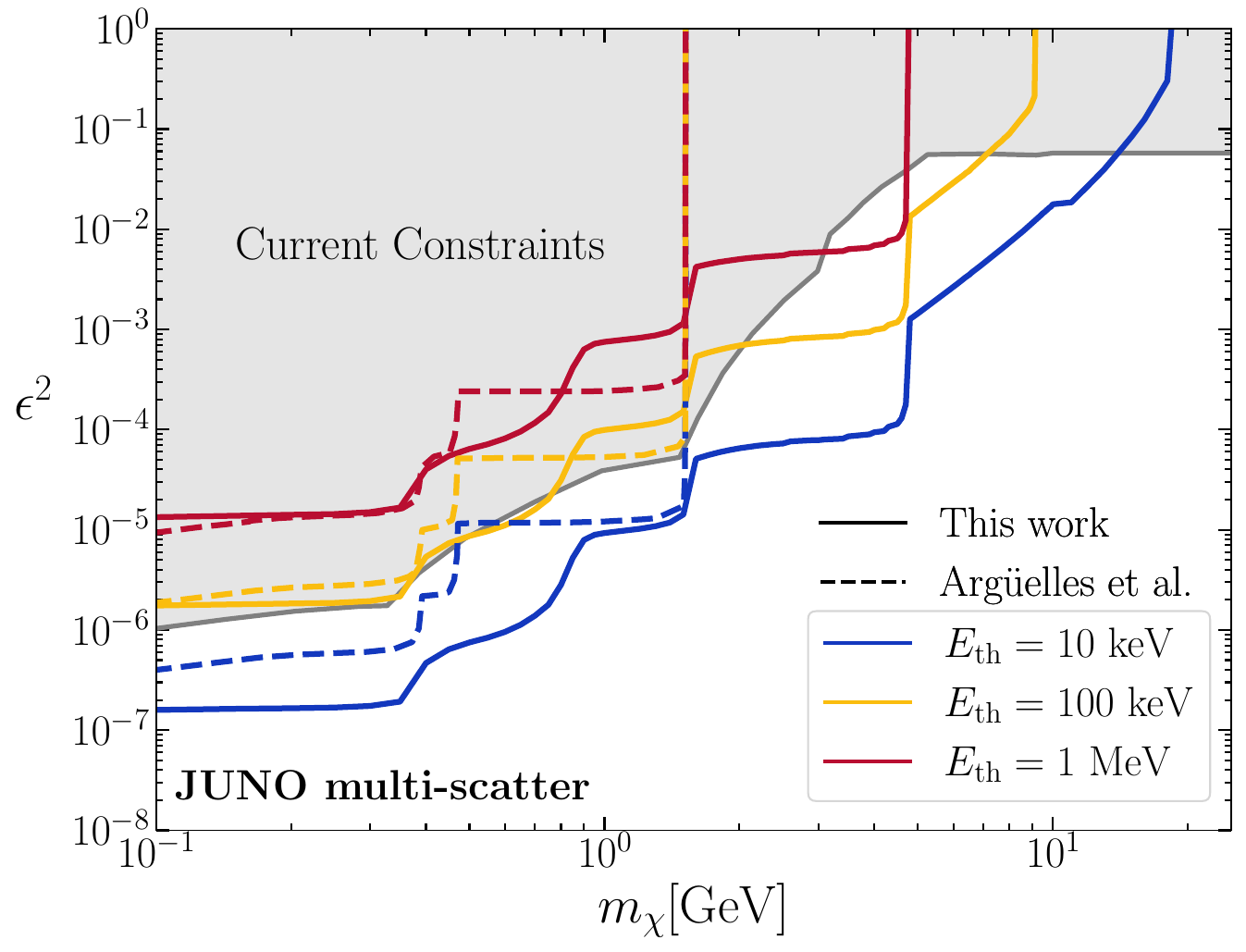}    
    \caption{Comparison of our results for MCP limits and projections with those of previous work in~\cite{ArguellesDelgado:2021lek}. {\it Left:} The single scatter constraints from SuperK and sensitivity with JUNO. {\it Right:} Multiple scatter sensitivities with JUNO.}
    \label{fig:compareresult}
\end{figure*}

In Figure~\ref{fig:compareresult} we compare our results with the previous constraints and sensitivities derived in~\cite{ArguellesDelgado:2021lek} (which updated the original work of \cite{Plestid:2020kdm}) for both single and multiple scatter searches. 

At $m_\chi\lesssim$~GeV our limits and projections for the single-scatter and the $E_{\rm th}=10$~keV multi-scatter cases are stronger than those in~\cite{ArguellesDelgado:2021lek}. This is expected due to our inclusion of production from proton bremsstrahlung, and we have checked that there is good agreement when we remove bremsstrahlung from our results. The sensitivities at GeV~$\lesssim m_\chi\lesssim 1.5$~GeV, where $J/\psi$ decay dominates, are similar.  

Meanwhile for multi-scatter signals with $E_{\rm th}=100$~keV and $E_{\rm th}=1$~MeV there are differences between our results and those of Ref.~\cite{ArguellesDelgado:2021lek} even when the $J/\psi$ decay dominates. We attribute these differences to differing choices in how we compute the multiple scattering rates at relatively high thresholds. In particular, in Eq.~\eqref{eq:Nmulti} we evaluate $N_{\rm single}$ and $P_1(E_{r,{\rm min}})$ with the value of $E_{r,\min}$ used in determining $P_{n\geq 2}(E_{r,{\rm min}})$, which appears to differ from what is done in Ref.~\cite{ArguellesDelgado:2021lek}. Our approach might give a better estimate of the multi-scatter rate, however a dedicated energy-dependent multi-scatter Monte Carlo analysis will be required to obtain fully reliable projections.

Finally, our inclusion of $\Upsilon$ decays and the Drell-Yan process means that (unlike Ref.~\cite{ArguellesDelgado:2021lek}) we find sensitivities that extend into the range $m_\chi\gtrsim 1.5$~GeV.

\bibliography{MCP}

\end{document}